\documentclass[preprint, 12pt, 3p, numbers, authoryear]{elsarticle}

\usepackage{lipsum} 	
\usepackage{graphicx}	
\usepackage{float}		

\usepackage{booktabs}
\usepackage[normalem]{ulem}
\useunder{\uline}{\ul}{}

\usepackage{longtable,booktabs}
\usepackage{multirow}
\usepackage[table,xcdraw]{xcolor}
\usepackage{mathtools}
\usepackage{amssymb}
\usepackage{amsmath}
\usepackage{caption}
\usepackage{subcaption}
\usepackage{tikz}

\usepackage[stable]{footmisc}

\usepackage{arydshln} 

\usepackage{framed} 
\usepackage{nomencl} 
\makenomenclature
\usepackage{etoolbox}
\renewcommand\nomgroup[1]{%
  \item[\bfseries
  \ifstrequal{#1}{L}{List of Symbols}{%
  \ifstrequal{#1}{P}{Superscripts}{%
  \ifstrequal{#1}{B}{Subscripts}{}}}%
]}

\makeatletter
\newcommand{\vast}{\bBigg@{4}}
\newcommand{\Vast}{\bBigg@{7}}
\makeatother

\usepackage{fancyhdr}
\makeatletter
\def\ps@pprintTitle{%
	\let\@oddhead\@empty
	\let\@evenhead\@empty
	\def\@oddfoot{\reset@font\hfil\thepage\hfil}
	\let\@evenfoot\@oddfoot
}
\makeatother

\usepackage{framed} 
\usepackage{nomencl} 
\makenomenclature
\usepackage{etoolbox}
\renewcommand\nomgroup[1]{%
  \item[\bfseries
  \ifstrequal{#1}{L}{List of Symbols}{%
  \ifstrequal{#1}{C}{Superscripts}{%
  \ifstrequal{#1}{B}{Subscripts}{}}}%
]}

\journal{Atomization \& Sprays}
\begin{document}

\begin{frontmatter}

\title{Detailed numerical simulations of atomization of a liquid jet in a swirling gas crossflow}

\author[label1,label2,label5]{Surya Prakash R}

\author[label1,label4]{Suhas S Jain}

\author[label3]{Jeffery A. Lovett}

\author[label2]{B N Raghunandan}

\author[label1]{R V Ravikrishna}

\author[label1]{Gaurav Tomar \corref{cor1}}
\ead{gtom@iisc.ac.in}
\cortext[cor1]{Corresponding author}

\address[label1]{Department of Mechanical Engineering, Indian Institute of Science, India}
\address[label2]{Department of Aerospace Engineering, Indian Institute of Science, India}
\address[label3]{Pratt \& Whitney Aircraft Engines, East Hartford, USA}
\address[label4]{Department of Mechanical Engineering, Stanford University, USA}
\address[label5]{Department of Mechanical Engineering, Indian Institute of Technology Dharwad, India}

\begin{abstract}
Breakup of a liquid jet in a high speed gaseous crossflow finds wide range of engineering and technological applications, especially in the combustors of the gas turbine engines in aerospace industry. In this study, we present volume-of-fluid method based direct numerical simulations of a liquid jet injected into a swirling crossflow of gas. The liquid is injected radially outwards from a central tube to a confined annular space with a swirling gas crossflow. The essential features of the jet breakup involving jet flattening, surface waves and stripping of droplets from the edges of the jet are captured in the simulations. We discuss the effect of swirl on the spray characteristics such as jet trajectory, column breakup-length, and size, shape-factor and velocity distribution of the drops. Drop size increases with swirl and penetration is slightly reduced. Moreover, the trajectory follows an angle (azimuthal) that is smaller than the geometric angle of the swirl at the inlet.  Interestingly, we also observe coalescence events  downstream of the jet that affect the final droplet size distribution for the geometry considered in this study.

\end{abstract}

\begin{keyword}
atomization and sprays \sep liquid jet in crossflow \sep swirling flow \sep column-breakup length \& drop characteristics 
\end{keyword}

\end{frontmatter}

\newpage

\pagenumbering{arabic}
\setcounter{page}{1}

\section{Introduction}

Liquid jets in subsonic gaseous crossflow (LJICF) finds applications in various industries including agricultural sprays  and in the aerospace industry in the combustors of gas-turbine engines, afterburners, ramjets and scramjets. With the aim of reducing the pollutant emissions by increasing the efficiency of combustion, there has been a renewed interest in this field over the past decade. There are multiple ways to increase the efficiency of combustion in LJICF systems, one of them being the use of swirling-gaseous crossflow. The swirling gas flow increases the turbulence intensity in the air stream, decreases the residence time, thus ensuring better mixing of the gas with the fuel and therefore leads to enhanced combustion efficiency. 






A number of experimental studies have been performed on LJICF over the years \citep{wu1997breakup,wu1998spray,dhanuka2011lean}. Most of these studies discuss experiments at near-atmospheric conditions due to the obvious difficulty in setting up the experiments at high pressure and temperature conditions. Thus, a numerical investigation can complement the experimental observations to attain a better understanding of the involved physics.   
Due to the constraints on experimental conditions and choice of the working liquid, previous experimental studies on LJICF were mostly limited to high density ratio flows \citep{vich1997destabilisation,mazallon1999primary,wu1997breakup}, whereas computational studies have been performed for both low and high density ratios \citep{aalburg2005properties,elshamy2005study,elshamy2007experimental,herrmann2010detailed,herrmann2011influence,behzad2015azimuthal,behzad2016surface,xiao2013large,li2014high}. A comparative study for density ratios of 10 and 100 by Hermann\citep{herrmann2011influence} showed that with an increase in the density ratio: (a) penetration of the liquid jet increases, (b) bending of the liquid jet reduces, (c) spread of the spray in the transverse direction increases and (d) wavelength of the traveling waves on the jet and the surface waves seen on the jet decreases, thus producing smaller droplets. In our recent study on secondary breakup of liquid drops \citep{jain2019secondary,jain2018a}, we also showed that the density ratio is an important parameter characterizing the secondary breakup of drops.   



 
There is abundant experimental literature that quantifies the spray characteristics such as spray trajectory \citep[see,][]{geery1969penetration,kush1973liquid,horn1971investigation,schetz1977penetration,schetz1980wave,nejad1983effects,nejad1984effects,less1986transient,kitamura1976stability,nguyen1992liquid,karagozian1986analytical,higuera1993incompressible,amighi2009trajectory} and jet penetration \citep{birouk2007role,thawley2008evaluation,stenzler2006penetration,wu1997breakup} at ambient atmospheric conditions and in non-atmospheric conditions \citep{bellofiore2007air,elshamy2005study,elshamy2007experimental}. Overall, the statistical behavior of the sprays has been well established from the copious experimental studies. However, numerical studies do not predict the jet trajectory accurately  \citep[see,][]{herrmann2010detailed,xiao2013large}. This difference between the computed and the experimentally observed trajectories has been attributed to the lack of applicability of the jet trajectory correlations obtained at normal pressures to high pressure conditions, that is at low density ratios $< 100$ \citep{cavaliere2003bending}.






Breakup of liquid jets has been suggested to be analogous to the secondary breakup of drops. Similar to drops, jets have been found to display regimes of bag, multimode and shear breakup that are governed by Weber number \citep{vich1997destabilisation,mazallon1999primary,wu1997breakup}. Although there is no good agreement on the dependence of the nature and the location of the breakup of a jet on the momentum-flux ratio and the Weber number \citep{lee2007primary,wu1997breakup,bellofiore2007air}, broadly two types of breakups have been observed that result from the disturbances on the jet windward surface: (a) surface breakup, which corresponds to the separation of ligaments and droplets from the jet surface close to the injector. This breakup is thought to be due to the azimuthal disturbances that lead to the formation of the interface corrugations and eventually into ligaments and drops \citep{xiao2013large,behzad2015azimuthal,behzad2016surface} or a ``boundary-layer stripping mechanism" wherein the liquid drops are pinched-off  from the viscous boundary layer that forms at the jet periphery due to the shear between the gas and the liquid phases \citep{sallam2004breakup}, (b) column breakup, which refers to the breakup of the liquid column as a whole due to the axial disturbances. This breakup is thought to be due to either Rayleigh-Taylor instability \citep{mazallon1999primary,sallam2004breakup,ng2008bag,xiao2013large} or Kelvin-Helmholtz instability \citep{schetz1980wave}. These two breakup mechanisms constitute the primary atomization of the liquid jet. However, the ligaments and droplets formed from these may undergo further breakup leading to the formation of finer droplets (a liquid spray), also known as secondary atomization.

Recent developments of numerical algorithms for simulations of two-phase flows have made highly resolved simulations of atomization processes possible.
\cite{herrmann2010detailed,herrmann2011influence} performed highly resolved simulations of breakup of a liquid jet at low density ratios. \cite{xiao2013large} performed high density ratio simulations of breakup of high speed turbulent jets and used the experimental results from \cite{sallam2004breakup} and \cite{elshamy2007experimental} for comparison of laminar and turbulent jets. \cite{li2014high} performed simulations of breakup of swirling liquid jets. Although simulations of direct injection jets and swirling jets and that of breakup of low speed liquid jet by a crossflow have been performed, breakup  of a liquid jet in a swirling crossflow has not been studied in detail. Detailed analysis such as drop-size distribution and drop-shape distribution are still lacking. Our preliminary experimental study on the effect of swirling crossflow on the atomization characteristics suggests an increase in the sauter mean diameter (SMD) for the geometry considered and an interesting linear correlation for the bending of the jet trajectory in the azimuthal direction \cite{prakash2017e}. In the present study, we investigate the effect of swirl on the breakup of a liquid jet in crossflow. We present the evolution of the spray characteristics and correlations for jet trajectory. We also discuss the evolution of the shape-factor distribution.

The paper is organized as following: (a) section \ref{sec:problem} describes the problem statement along with the parameters used, (b) section \ref{sec:equation} describes the governing equations and the numerical method used in the current study, (c) section \ref{sec:flow} describes the interface structures observed in the flow along with the breakup mechanisms, (d) section \ref{sec:traj} presents the trajectory correlation for all the cases in the study, (e) in section \ref{sec:distribution}, we discuss the drop-size, shape-factor and drop-velocity distributions and (f) finally, in section \ref{sec:conclusion}, we summarize the results along with the concluding remarks.

\section{Problem description\label{sec:problem}}

\begin{figure}[t!]
\centering
\includegraphics[width=0.65\textwidth]{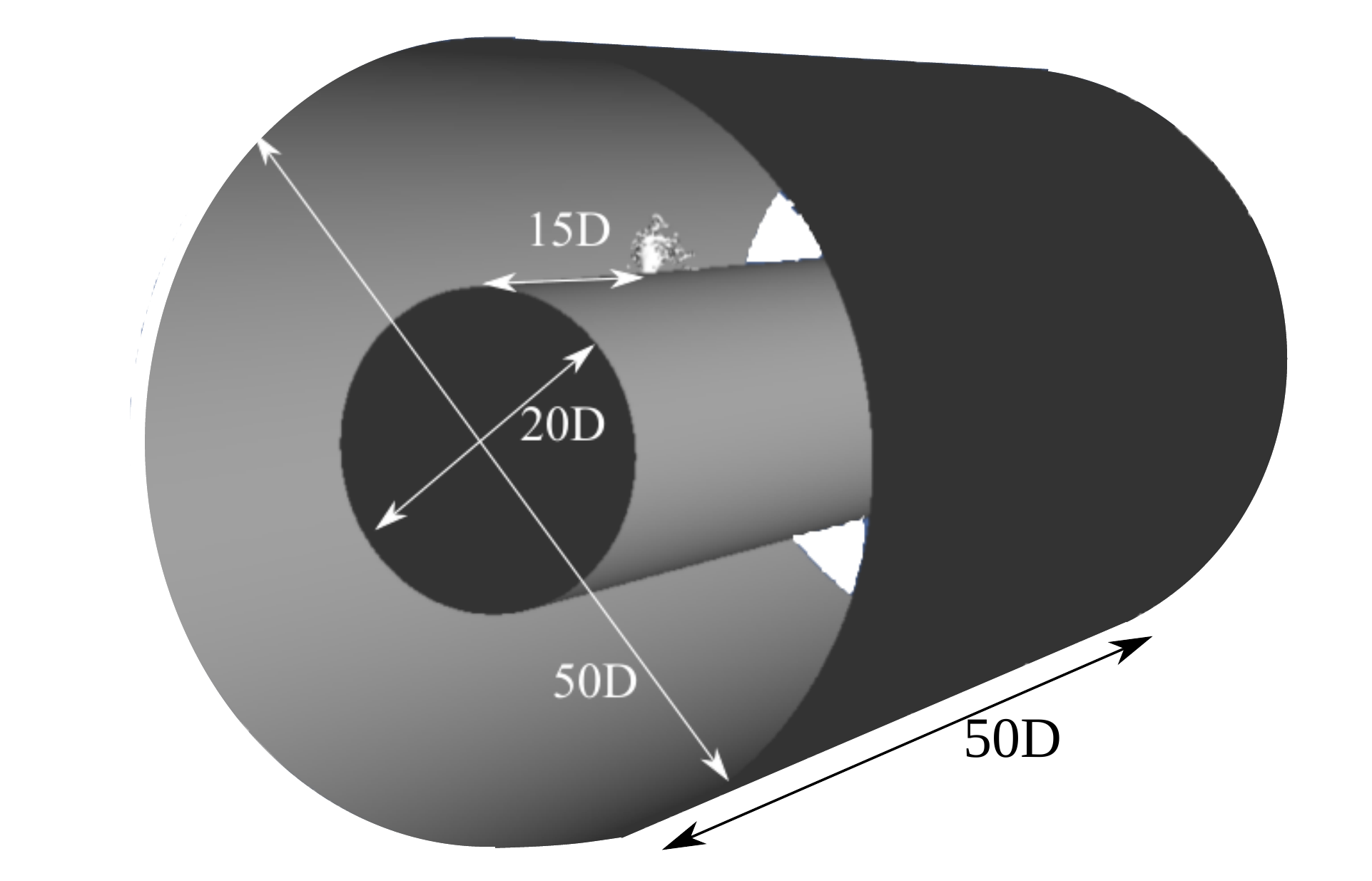}
\centering
\caption{Domain along with the dimensions in terms of the jet diameter $D$. Jet is shown in white color.}
\label{fig:domain}
\end{figure}
In the present study, we simulate a liquid jet injected into a swirling-gas crossflow (LJISCF). Most of the previous numerical studies simulate jet in a rectangular domain and focus their investigation only on the regions very close to the jet to study the disintegration of the liquid column. However, to completely understand the complex atomization process and to develop models for realistic scenarios, a full simulation of the liquid column and the resultant spray is necessary. Hence we use a realistic annular domain to study the breakup of a liquid jet injected into gas crossflow using a long computational domain of $35D$ from the point of injection and adaptively refine the grid to capture very fine droplets formed in the flow, where $D$ is the diameter of the jet at the injector. We choose to study two liquid-to-gas momentum-flux ratios (20 and 25 and we represent them as q20 and q25 in the rest of this article) and study the flow behavior and breakup patterns in the flow. Figure \ref{fig:domain} shows the computational domain which is an annular region with the ratio of inner to outer radii 2:5 chosen to accommodate the essential features of the swirling-gas flow such as a few rotations of the flow along the axial direction. Length along the axis of the domain is $50D$. Jet is located at $15D$ from the inlet of the gas and the exit of the spray is located at $35D$ from the injection point of the jet. This length of the domain was selected based on the preliminary analysis of sphericity of the drops using a domain length of $100D$ on a coarser grid. This preliminary study suggested that the primary atomization is complete before 30D from the point of injection. The jet diameter is chosen to be $1mm$. The liquid and gas densities are represented by $\rho_l$ and $\rho_g$, respectively, and viscosities by $\mu_l$ and $\mu_g$, respectively. The crossflow velocity is $U_g$ and the liquid jet inlet velocity is $U_l$. 

The non-dimensional parameters that govern the flow can be expressed in terms of the non-dimensional numbers, namely, aerodynamic/gas Weber number $We_{g}=\rho_g U_g^2 D/\sigma$, liquid Weber number $We_l=\rho_l U_l^2 D/\sigma$, liquid-to-gas momentum-flux ratio $Q=\rho_l U_l^2/\rho_gU_g^2$, swirl number $SN=G_{\theta}/RG_x$ of the gas flow (defined as the ratio of axial flux of the tangential momentum to the axial flux of the axial momentum), Reynolds number $Re_g=\rho_g U_g D/\mu_g$ of the gas flow and liquid Reynolds number $Re_l=\rho_l U_l D/\mu_l$, where $G_{\theta}=\int_{R1}^{R2}\rho_g u v r^2 dr$, $G_x=\int_{R1}^{R2}\rho_g u^2 r dr$, $R=R2-R1$, $R1$ and $R2$ are the two radii of the annulus region. Here, $u$ is the axial component of the gas velocity and $v$ is the azimuthal component. We note here that the axial velocity in these simulations is kept constant and increase in the swirl of the flow increases the absolute velocity. Values of the non-dimensional numbers and other parameters used in the simulations are listed in Table \ref{tab:sim}. Density ratio of $\rho^*=180$ corresponds to an elevated pressure of $5\mathrm{bar}$ in the annular region which is typically observed in gas-turbine engines. 
Numerical investigations are carried out for various values of swirl number and liquid-to-gas momentum-flux ratio such as $SN=0, 0.42$ and $0.84$ and $Q=20$ and $25$ and are listed in Table \ref{tab:case}. Hereafter, we refer to these cases as listed in  Table \ref{tab:case}. 

\begin{table}[H]
\centering
\caption{Parameters used in the simulation.}
\label{tab:sim}
\begin{tabular}{cc}
\hline
parameter & values \\ \hline

gas density $\rho_g$ & $1$\\[0.1cm]
liquid density $\rho_l$ & $180$\\[0.1cm]
liquid-to-gas density ratio $\rho^*$ & $180$\\[0.1cm]
gas viscosity $\mu_g$ & $1.3\times10^{-5}$\\[0.1cm]
liquid viscosity $\mu_l$ & $1.32\times10^{-3}$\\[0.1cm]
liquid-to-gas viscosity ratio $\mu^*$ & $101.538$ \\[0.1cm]
surface tension $\sigma$ & $0.00267$\\[0.1cm]
gas Reynolds number $Re_g$ & 9230.77\\[0.1cm]
liquid Reynolds number $Re_l$ & 5454.55\\[0.1cm]
gas Weber number $We_g$ & 134.83\\[0.1cm]
liquid Weber number $We_l$ & 2696.63\\[0.1cm]
\hline
\end{tabular}
\end{table}

\begin{table}[H]
\centering
\caption{Test cases.}
\label{tab:case}
\begin{tabular}{c|cc}
\hline
parameter & $Q=20$ & $Q=25$ \\[0.1cm] \hline

$SN=0$  	&	q20sn0		&  	q25sn0		\\[0.1cm] 
$SN=0.42$ 	&	q20sn42		&	q25sn42		\\[0.1cm]
$SN=0.84$	&	q20sn84		&	q25sn84		\\

\hline
\end{tabular}
\end{table}

 Maintaining other non-dimensional parameters, the ideal analytical velocity profiles are utilized in the simulations to obtain the swirling flow in the annular space. Gas velocity used in the simulations $U_g=3(\hat{i} + SN z\hat{j} - SN y\hat{k})$. Here, $\hat{i}$ is in the axial direction, and $\hat{j}$ and $\hat{k}$ are orthonormal cartesian coordinates in the plane perpendicular to the axial direction. A ramp profile for the gas velocity given by $U_g(t)=U_gtanh(t/\tau)$ is used at the start of the simulation to stabilize the numerical simulations, where $\tau$ is an appropriate time constant, chosen to be roughly $1/10^{th}$ of one flow-through time. The liquid jet is injected radially outwards from the inner annular tube having a plug velocity profile, assuming the nozzle to have a very small length-to-diameter ratio. Liquid velocities used in the simulations are $U_l=1\hat{j}$ for q20 and $U_l=1.12\hat{j}$ for the q25 case. All the remaining walls in the domain are given a no-slip boundary condition.

\section{Governing Equations\label{sec:equation}}

Assuming both liquid and surrounding gas to be incompressible, the continuity equation is given by,

\begin{equation}
\nabla \cdot \vec{u}=0
\end{equation}

where $\vec{u}$ is the velocity field. Now, dividing the domain into two subdomains for each of the phases, Navier-Stokes equations hold in each of the subdomains and govern the conservation of momentum in each of the phases, and the fluid motion in the two subdomains are coupled at the interface with the following jump conditions:
\begin{equation}
    [\vec{u}]_s = 0,
    \label{equ:velj}
\end{equation}
\begin{equation}
    -[-p + 2\mu\vec{n}\cdot\textbf{S}\cdot\vec{n}]_s = \sigma \kappa,
    \label{equ:normj}
\end{equation}
\begin{equation}
    -[2\mu \vec{t}\cdot\textbf{S}\cdot\vec{n}]_s = \vec{t}\cdot\vec{\nabla}_s\sigma,
    \label{equ:tanj}
\end{equation}
where, $[x]_s=x_2 - x_1$ denotes \textit{jump} across the interface and $x_1$ and $x_2$ are the values of the quantity $x$ on the two sides of the interface, $p$ is the pressure, $\mu$ is the viscosity, $\vec{n}$ is the local unit normal at the interface, $\textbf{S}=[(\nabla \vec{u})+(\nabla \vec{u})^{T}]/2$ is the deformation rate tensor, $\sigma$ is the surface tension coefficient, $\kappa$ is the local mean curvature of the interface, $\vec{t}$ represents the local unit tangent vector at the interface, and $\vec{\nabla}_s$ represents the surface gradient. For a more detailed discussion on the equations that govern the fluid flows with interfaces and the corresponding jump conditions, see Section 2.4 in \citet{tryggvason2011direct}. One approach to solve the above coupled two-fluid flow problem in two subdomains is with the use of ``\textit{one-fluid model}" \citep{kataoka1986local,tryggvason2011direct,Mirjalili2017}, where the global momentum balance is given by the Navier-Stokes equations augmented with surface forces to implicitly account for the interfacial jump conditions, namely, continuity of velocity (Equation \ref{equ:velj}), and normal stress balance (Equation \ref{equ:normj}) and tangential stress balance (Equation \ref{equ:tanj}). These equations, using the one-fluid formulation, can be written as:
\begin{equation}
\rho\Big[\frac{\partial\vec{u}}{\partial t}+\nabla \cdot (\vec{u}\vec{u})\Big]=-\nabla p + \nabla \cdot (2\mu \textbf{S}) + \sigma \kappa \vec{n} \delta_{s},
\end{equation}
where the density $\rho=\rho_{l}F+\rho_{g}(1-F)$, $F$ is the volume fraction of liquid that takes values between 0 and 1, and $\rho_{l}$ and $\rho_{g}$ are liquid and gas densities, respectively. Similarly, $\mu$ is the viscosity and can be expressed as $\mu=\mu_{l}F+\mu_{g}(1-F)$, where $\mu_{l}$ and $ \mu_{g} $ are liquid and gas viscosities, respectively. The last term in the equation accounts for the surface tension force at the interface embedded in an Eulerian grid: $f_s=\sigma \kappa \vec{n}$, where $\delta_{s}$ is the surface Dirac-delta function. The surface tension force is modeled as a volumetric force using the continuum surface force approach by \cite{brackbill1992continuum}. The evolution equation for the interface is given as an advection equation in terms of the volume fraction, $F$,
\begin{equation}
\frac{\partial F}{\partial t} + \vec{u} \cdot\nabla F = 0.
\end{equation}
For simulations of the flow using these governing equations, we use Gerris solver that uses a geometric volume-of-fluid (GVOF) method on a cell-based-octree-AMR (adaptive mesh refinement) grid \citep[see,][]{popinet2003gerris,popinet2009accurate,tomar2010multiscale}. Gerris uses a second-order accurate staggered-time discretization for velocity, volume-fraction and pressure fields. Balanced-force algorithm \citep{francois2006balanced} is used to calculate the surface tension forces. The discretized equations can be written as,
\begin{equation}
\rho_{n+\frac{1}{2}} \Bigg[\frac{\vec{u}_{\star}-\vec{u}_{n}}{\Delta t} + \vec{u}_{n+\frac{1}{2}} \cdot \nabla \vec{u}_{n+\frac{1}{2}}\Bigg] = \nabla \cdot[\mu_{n+\frac{1}{2}} (\textbf{S}_{n}+\textbf{S}_{\star})] + (\sigma \kappa \delta_{s} \vec{n})_{n+\frac{1}{2}},
\end{equation}
\begin{equation}
\frac{F_{n+\frac{1}{2}} - F_{n-\frac{1}{2}}}{\Delta t} + \nabla \cdot (F_{n} \vec{u}_{n})=0,
\label{equ:advection}
\end{equation}
\begin{equation}
\vec{u}_{n+1}=\vec{u}_{\star}-\frac{\Delta t }{\rho_{n+\frac{1}{2}}} \nabla p_{n+\frac{1}{2}},
\label{equ:pressure_corr}
\end{equation}
and 
\begin{equation}
\nabla \cdot\vec{u}_{n+1}=0.
\end{equation}
Here, the subscripts $(n+{1}/{2})$ define the intermediate time of the staggered time stepping adopted for the void-fraction field, velocity and density and the subscript $\star$ indicates the auxiliary-velocity field which is corrected using the pressure-correction equation (Equation \ref{equ:pressure_corr}), to obtain the velocity field at the next time step, $n+1$. Advection equation for $F$ (Equation \ref{equ:advection}) is solved in Gerris using geometric fluxing \cite{popinet2003gerris}. Adaptive mesh refinement is performed using a cost function based on the local vorticity in the field and the gradient of the void-fraction field, thus using a very fine refinement in the regions of high velocity gradient and at the interface. We use a thin transition region of two cells for smoothing the physical properties across the interface. To test the efficacy of the numerical algorithm in capturing high-density ratios, we have performed validation tests \citep{jain2015secondary}, which show good agreement with the analytical results. Additionally, the curved boundary of the domain due to the cylindrical annulus geometry is achieved in Gerris using a cut-cell method, where the boundary conditions are extrapolated onto the Cartesian mesh points \citep{popinet2003gerris}. 

\begin{figure}[t!]
\centering
\includegraphics[width=0.65\textwidth]{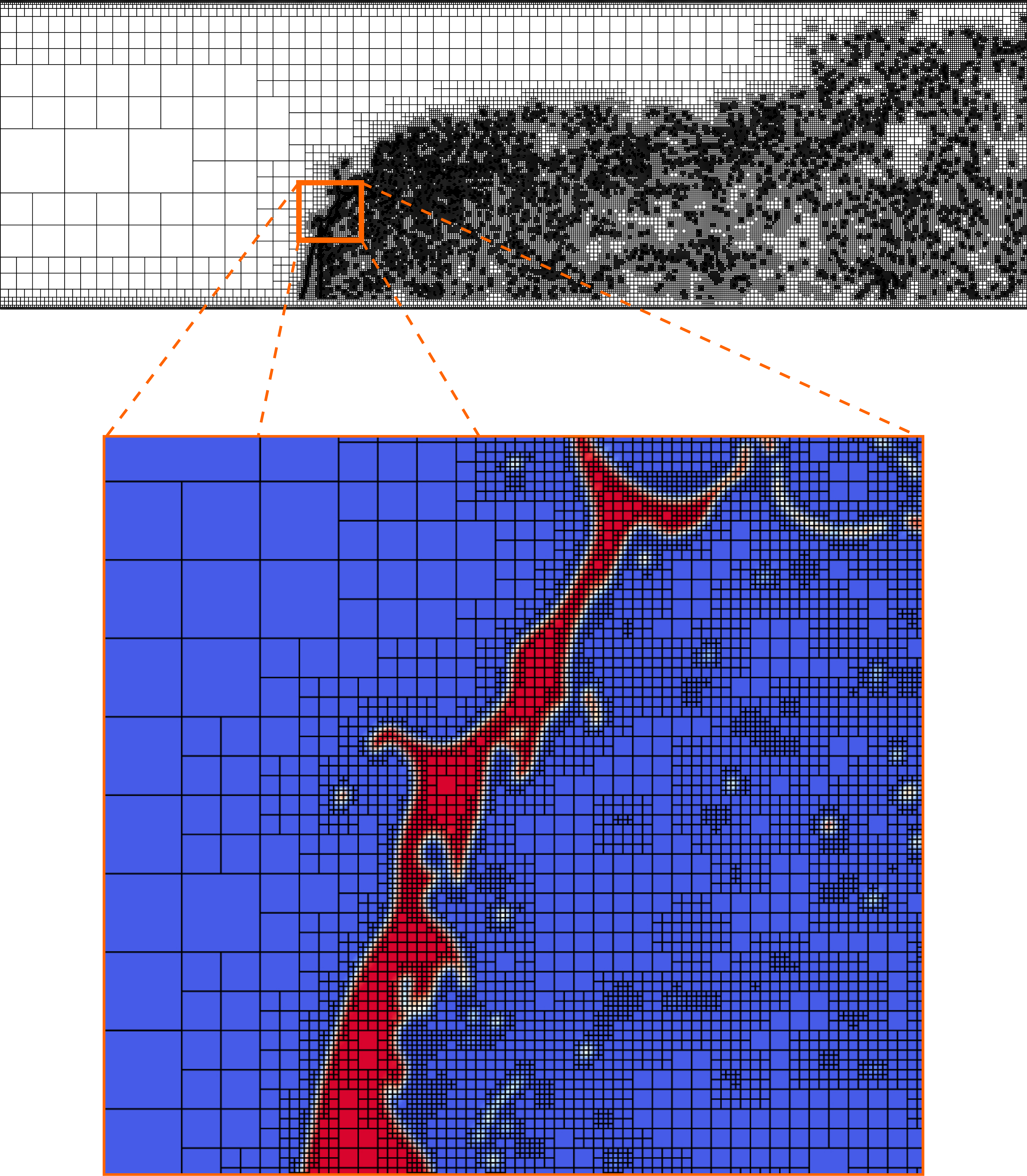}
\centering
\caption{A plane containing the annular axis and the injection point, showing the adaptive grid used in the simulation to illustrate various refined regions. A zoomed in view close to the jet is also shown, where red represents liquid region and blue represents the surrounding gas.}
\label{fig:grid_ref}
\end{figure}

Domain is discretized using a Cartesian AMR grid that is adaptively refined with a maximum refinement of 41 cells per diameter of the jet, $D$ (or $\approx24\mu m$ in physical space). A uniform grid with this resolution for the current domain size would have had a total of 5.28 billion cells. But with the use of AMR strategy, the number of cells has been reduced to $\approx7$ million. An estimate of the Kolmogorov scale in the crossflow is given by $\eta \sim D Re_g^{-\frac{3}{4}}=1.06\mu m$, and the estimate of the Hinze scale is given by $\sim \sigma/(\rho_g U^2)=7.42\mu m$. Though the smallest grid we use is  $\Delta\approx23 \eta$ and doesn't capture the detailed turbulent flow, we believe it is good enough to capture the breakup dynamics of the droplets produced by turbulent breakup, since $\Delta\approx3.3\eta_H$ and it has been shown  that the turbulent breakup of drops ceases at around $3-4.7\eta_H$ due to the balance of surface tension energy and turbulent kinetic energy \citep{martinez1999breakup}. However, the aerodynamic breakup is a more complex problem and might result in smaller droplets that cannot be resolved with any of the available current computational resources in a practical time without the use of subgrid models. Our simulations are also more resolved than most of the other recent studies (for example, \cite{li2014high} used a grid $\approx130$ times larger than $\eta$). In Appendix A we show that the mesh used for the simulations presented in this study shows convergence of the trajectory of the jet. 
  
Figure \ref{fig:grid_ref} shows the adaptively refined grid used in the simulations on a plane containing the annular axis and the injection point. Figure \ref{fig:grid_ref} also shows a zoomed in view of the jet illustrating the grid refinement used at various locations. Grid is most refined in the regions that contain higher gradient of the volume fraction $F$ and/or vorticity in the flow. A similar strategy has been used in previous studies of secondary breakup of drops \citep[see,][]{jain2015secondary,jain2019secondary}. A total of up to 128 cores were used for each of the simulations with an approximate simulation time of 30 days (an equivalent uniform grid would have taken 40 years with the same computational power), enough to flush the initial transients of the flow and to achieve a statistically stationary state.





\section{Flow structures and breakup mechanisms \footnote{A video of the jet breakup, presented at the Gallery of Atomization and Sprays - ICLASS 2018, can be found here: https://bit.ly/2Wjg42O}}\label{sec:flow}

\begin{figure}[t!]
\centering
\includegraphics[width=\textwidth]{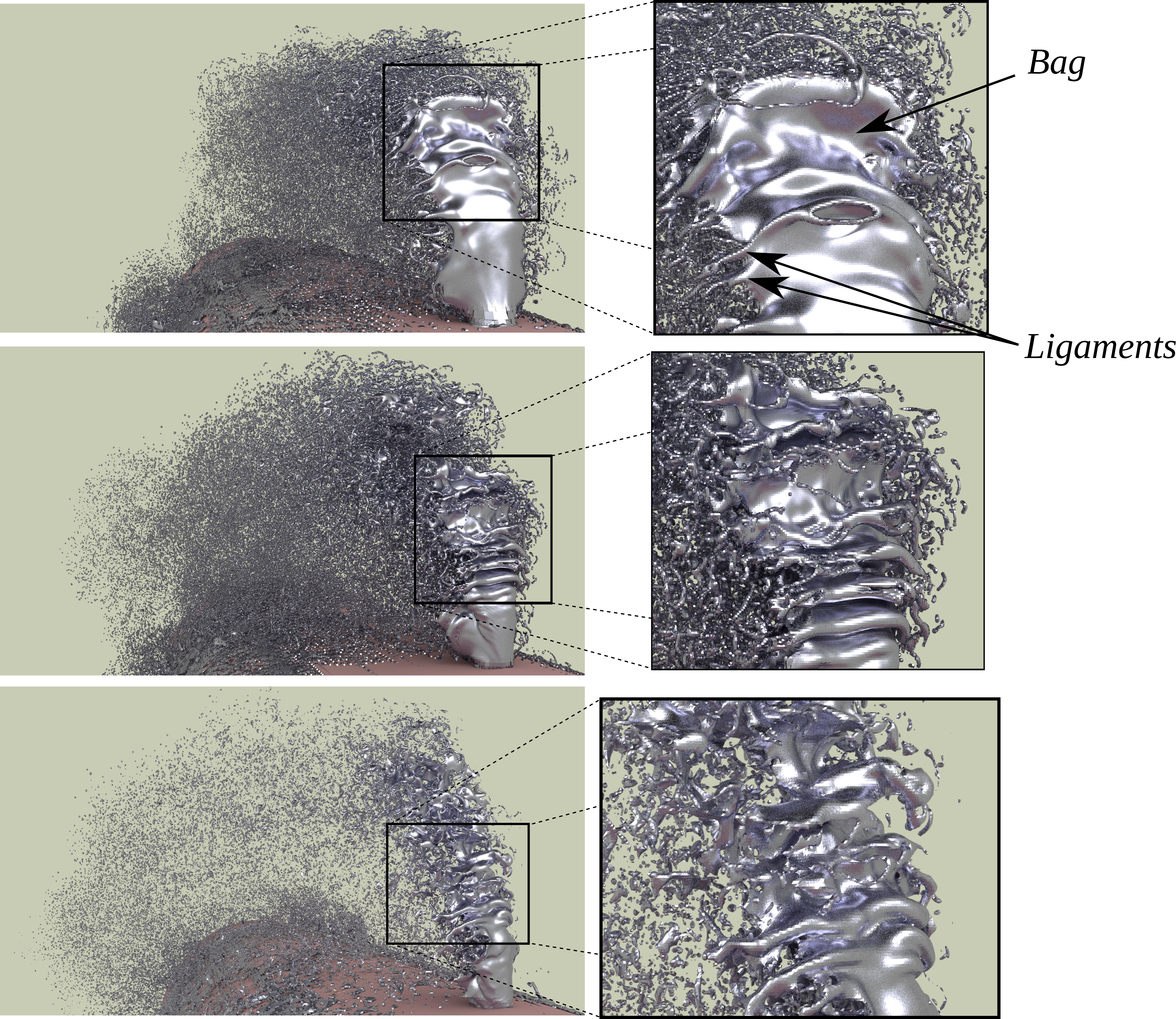}
\centering
\caption{Front view of the jet (in glossy grey) along with the inner cylinder (in pink) for cases from top to bottom (a) q20, sn0, (b) q20, sn42, (c) q20, sn84. Rendered using Blender (Ray-traced rendering package). Inset: Zoomed in image of the complex flow structures captured close to the location of primary atomization.}
\label{fig:render}
\end{figure}

\begin{figure}
    \centering
    \includegraphics[width=0.5\textwidth]{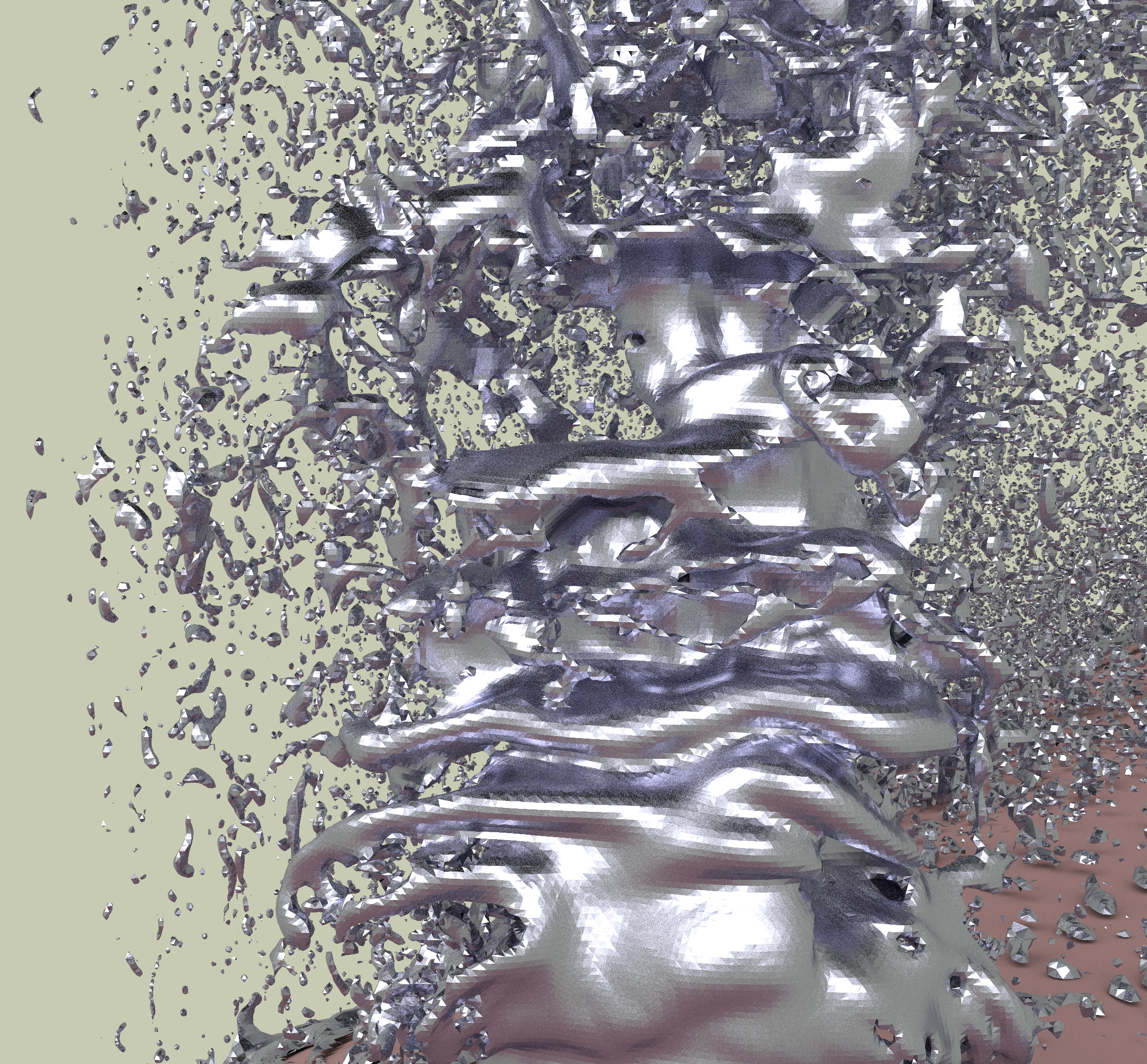}
    \caption{Zoomed in image of the windward jet surface from a view that is aligned along the swirling flow, showing no obvious bag-like structures for the case of q20, sn84.}
    \label{fig:renderfrontalign}
\end{figure}

Figure \ref{fig:render} shows a front view of the jet (in glossy grey) along with the inner cylinder (in pink) for cases from top to bottom (a) q20, sn0, (b) q20, sn42, (c) q20, sn84 (first column in the Table \ref{tab:case}. A zoomed in image of the jet close to the breakup point is also included to illustrate the complex flow structures captured in the simulations. It is clear from Figure \ref{fig:render} that the liquid jet column undergoes complex turbulent breakup for all the cases, though the liquid jet was laminar at the inlet. Breakup also appears to be more chaotic (with increasing small scales) and more turbulent with the addition of swirl in the crossflow and for increasing swirl numbers (sn42, sn84) compared to a non-swirling case (sn0). Zoomed in image also shows bag-like structures for sn0 case that is a characteristic of the column breakup process. However, bag like structures are not clearly seen for swirling cases and the jet windward surface is severely corrugated as shown in Figure \ref{fig:renderfrontalign} for sn84 case (looking along the direction of the swirling crossflow). A similar behavior of absence of bag structures for the turbulent jets was observed by \cite{xiao2013large}. Figure \ref{fig:render} also shows the formation of ligaments formed from the sides of the liquid column, as also observed previously in the experiments \citep{wu1997breakup,mazallon1999primary,sallam2004breakup}.  

\begin{figure}[t!]
\centering
\includegraphics[width=\textwidth]{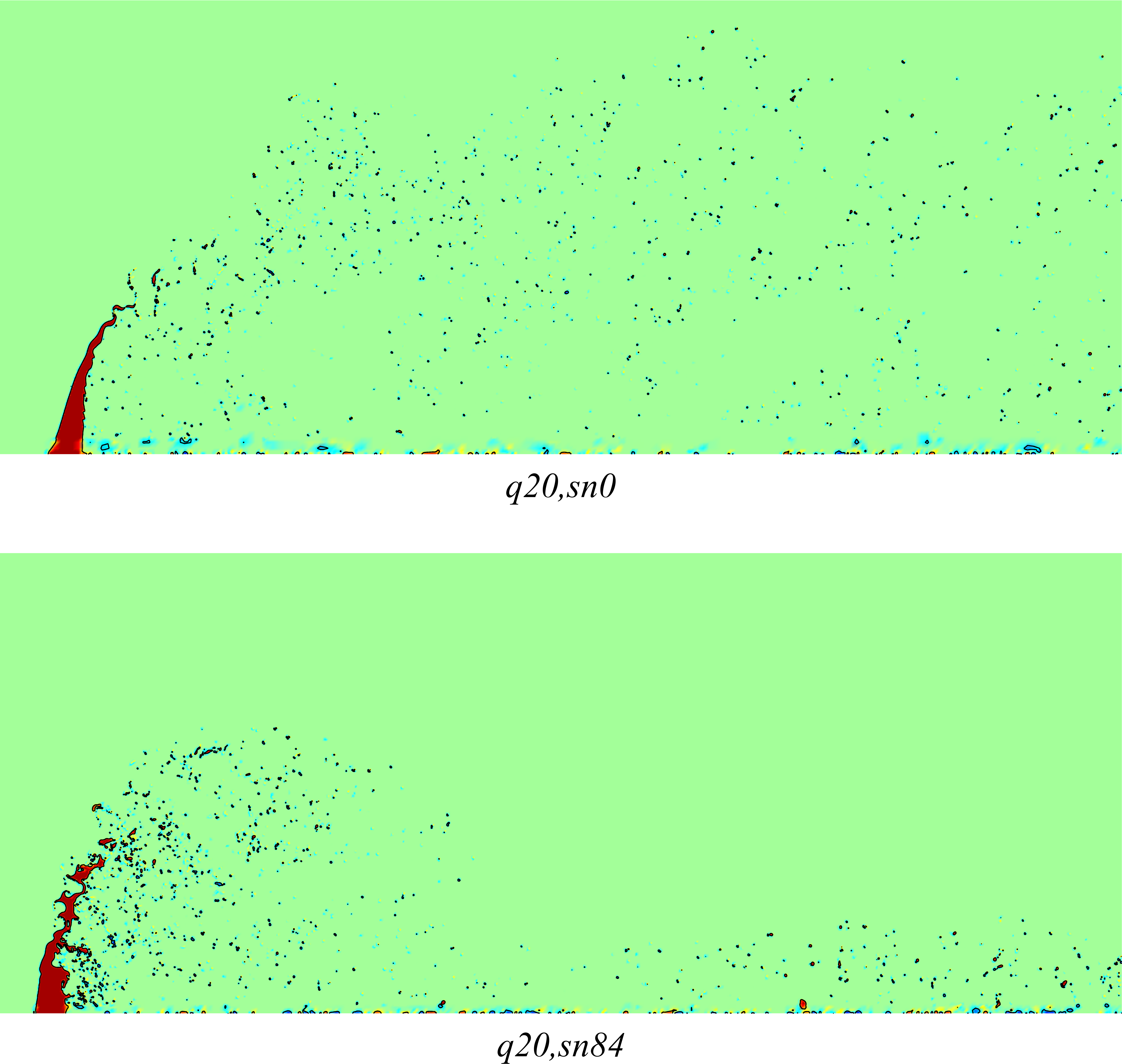}
\centering
\caption{Side view of the cross-section of the jet and spray along a plane passing through the annular axis and injection point for a swirling and non-swirling case. Top: q20, sn0. Bottom: q20, sn84.}
\label{fig:side_section}
\end{figure}

Figure \ref{fig:side_section} shows a side view of the cross-section of the jet and spray along a plane passing through the annular axis and injection point for a non-swirling case (q20, sn0) and a swirling case (q20,sn84). Two-dimensional axial waves can be seen on both the windward and leeward side of the jet for the q20sn0 case. A similar behavior was observed by \cite{behzad2016surface}. However, highly irregular surface of the jet on both windward and leeward sides can be seen for q20sn84 case. Figure \ref{fig:side_section} along with Fig.\ref{fig:render}, shows that the spray formed is in the axial plane for the non-swirling case and follows a helical path in the swirling case. 

\begin{figure}[t!]
\centering
\includegraphics[width=\textwidth]{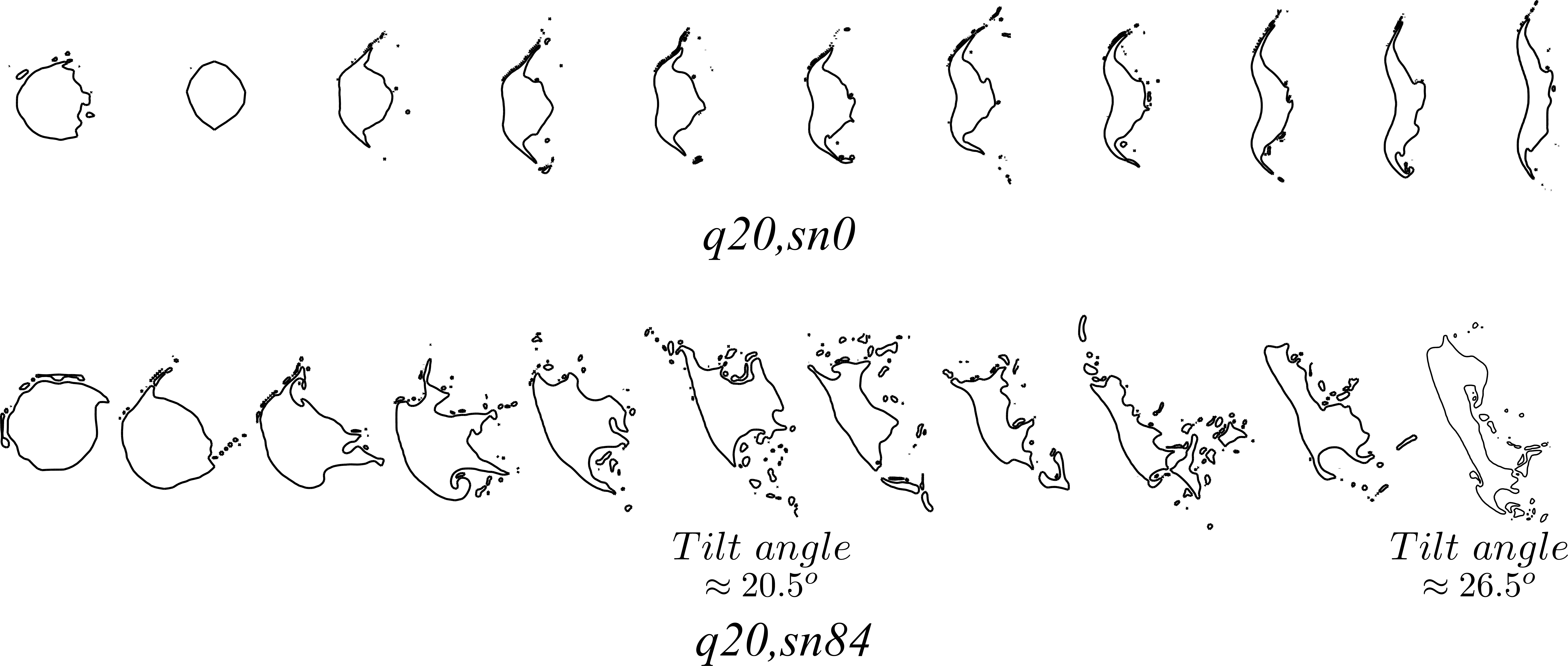}
\centering
\caption{Top view of the horizontal cross-section of the jet column at various heights separated by $0.5D$ from the point of injection for a swirling and non-swirling case. Top: q20, sn0. Bottom: q20, sn84.}
\label{fig:cross_section}
\end{figure}

Figure \ref{fig:cross_section} shows the top view of the horizontal cross-sections of the jet column at various heights separated by $0.5D$ from the point of injection for a non-swirling case (q20sn0) and a non-swirling case (q20,sn84). It clearly shows the sheet-thinning process on the sides of the jet for the q20sn0 case, which is a characteristic of the surface breakup of the jet. This process is very similar to the sheet-thinning process in the secondary breakup of drops \citep[see,][]{jain2015secondary,jain2019secondary}. Although, for q20-sn84 case, a surface breakup process that resembles sheet thinning is present, it is not symmetric about the mid plane (a plane containing the annular axis and the injection point).

\begin{figure}
    \centering
    \includegraphics[width=\textwidth]{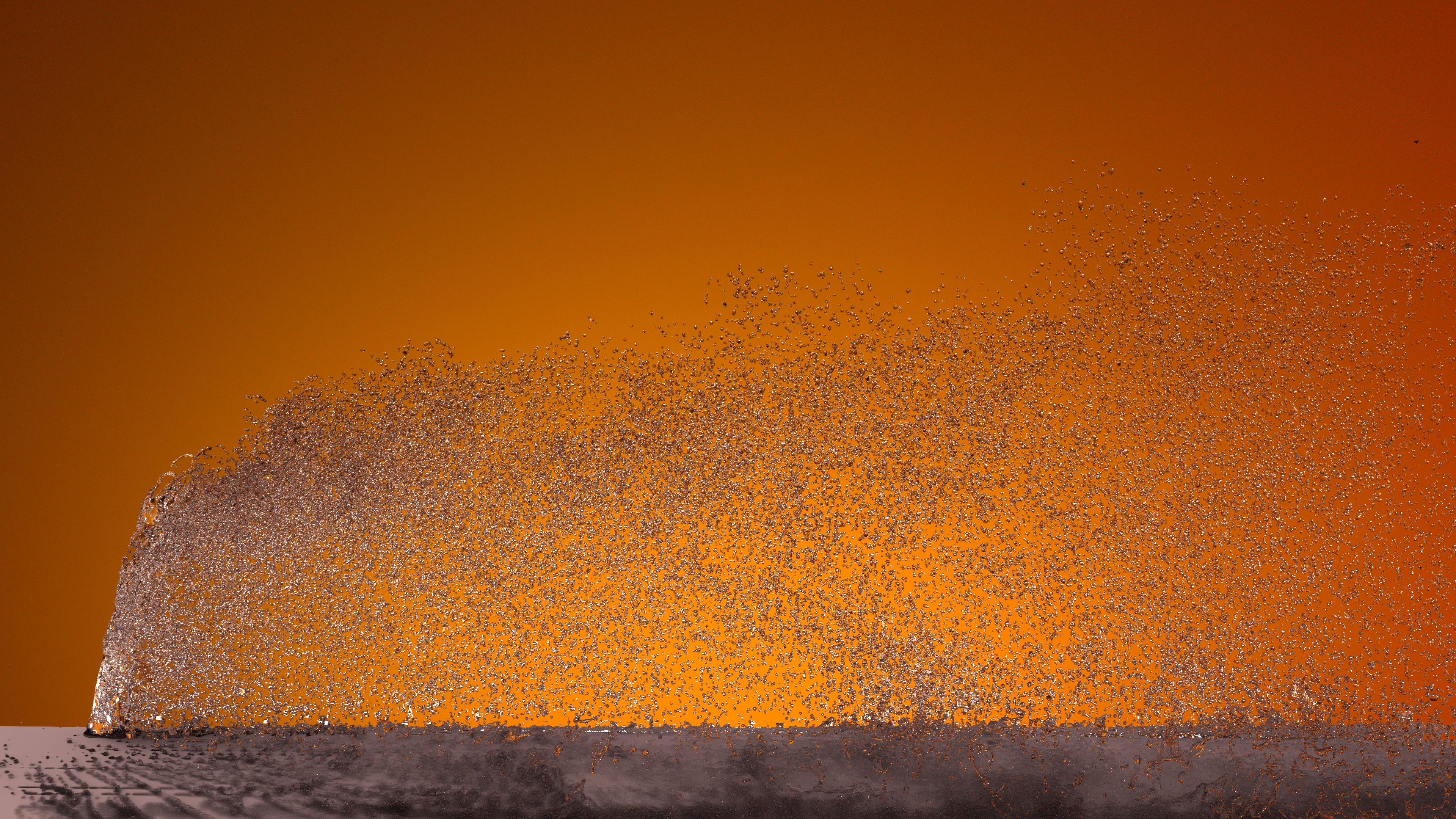}
    \caption{Side view of the fully developed jet and spray along with the inner cylinder (in pink) for the case q20, sn0. Rendered using Blender.}
    \label{fig:renderside}
\end{figure}

\begin{figure}[b!]
\centering

    \begin{subfigure}[b]{0.49\textwidth}
        \centering
       \includegraphics[width=\textwidth]{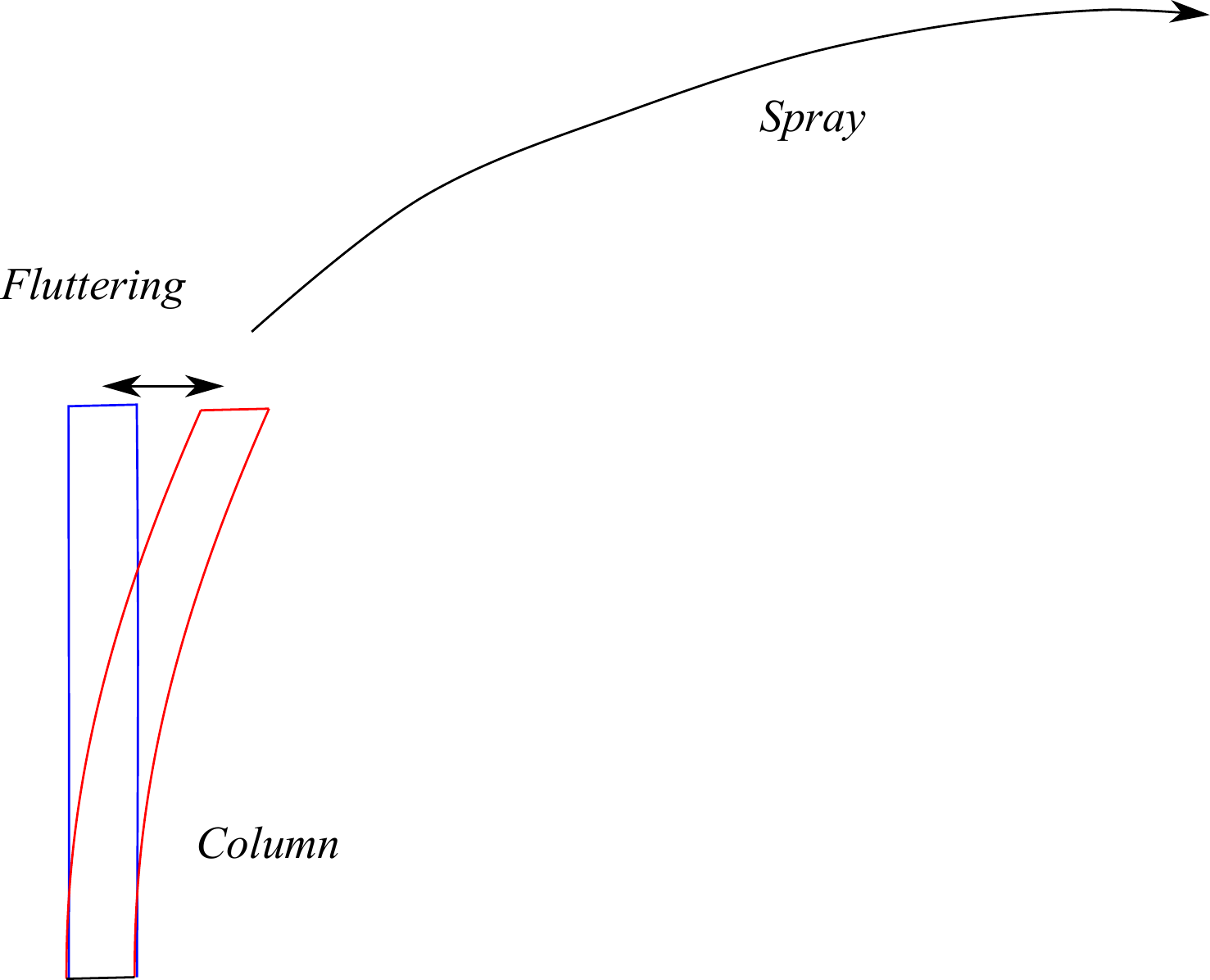}
        \caption{\label{fig:flutter} Schematic of column flutter.}
    \end{subfigure}%
    ~ 
    \begin{subfigure}[b]{0.49\textwidth}
        \centering
        \includegraphics[width=\textwidth]{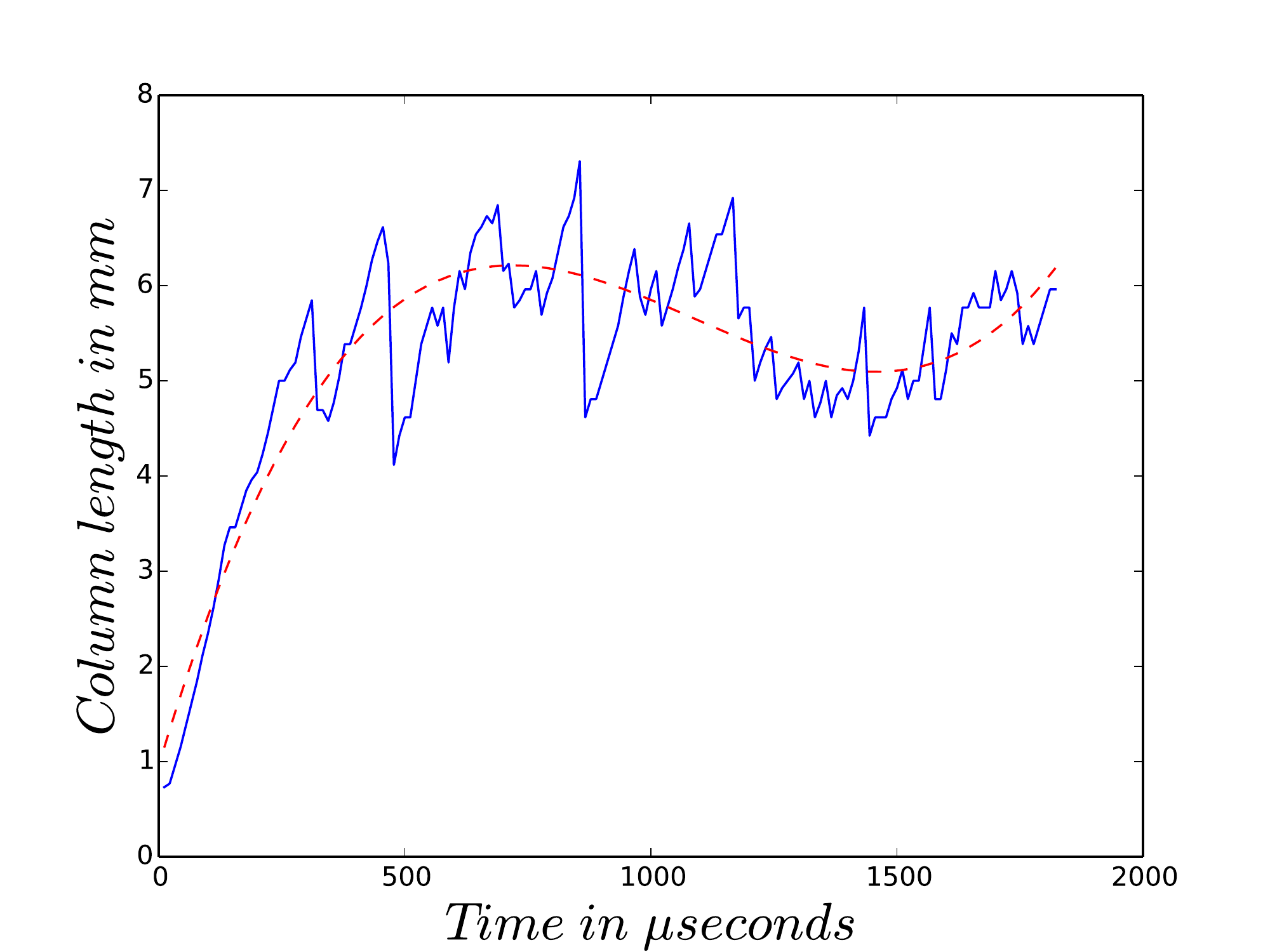}
        \caption{\label{fig:column} Column-breakup length for q10,sn0 (Blue line) along with the cubic polynomial fit (Red dotted line).}
    \end{subfigure}

\caption{Column flutter and column-breakup length.}
\label{fig:sn0column}
\end{figure}

The simulation is run for 4-5 flow-through time, and in this duration the spray has been observed to have reached a quasi-steady state after the initial disturbances caused by the issuance of the liquid jet tip into the crossflow environment has flushed completely. Figure \ref{fig:renderside} shows a realistic rendered snapshot of the fully developed jet and the spray from a side view. We define the quasi-steady state by examining the number of droplets passing through any given cross-section of the computational domain for a short duration of time. JICF is found to reach the quasi-steady state with some periodical variations occurring in the spray due to the whiplash action of the jet (due to the flutter in the column) (see Figure \ref{fig:flutter}). In the quasi-steady state, we found that there are around $4000\pm500$ droplets passing through any cross-section of thickness $2.5D$ for our configuration. We also estimated the characteristic frequency of the periodic variation in the spray (spray wave) and the whiplash action of the jet (column fluttering) and found that they agree well (Table \ref{tab:freq}), showing that the periodic variation in the spray could indeed be due to the column flutter/whiplash phenomenon. The characteristic frequency of the spray wave was found by fitting a sinusoidal curve onto the jet outer surface at various instantaneous time snapshots from the side view of the jet and by averaging the obtained frequency of the sinusoid over multiple time snapshots. The column flutter frequency was estimated by analyzing the time evolution of the back-and-forth motion of the tip of the jet column when viewed from the side.

Further to quantify the column breakup process, we compute the column length of the jet as a function of time and variation of the column length for the case of q20, sn0 in the transient period as shown in the Figure \ref{fig:sn0column}, where we define the column length as the intact length of the column or the height of the column from the point of injection where the first hole is formed in the liquid jet. A similar analysis is performed in the quasi-steady state to obtain the characteristic frequency associated with the column breakup length. Interestingly, we obtain two major breakup modes and the corresponding frequencies ($f_1,f_2$) are listed in the Table \ref{tab:freq}. Finally, we also compute the frequency of the surface waves on the column as demonstrated by Figure 8 in \cite{xiao2013large} for the non-swirling case (q20, sn0) and the characteristic frequency is listed in Table \ref{tab:freq}. We also observe that the frequency of the second mode of column-breakup length is close to the characteristic frequency of the surface waves on the column, suggesting that these two phenomenon could be correlated.   


\begin{table}[H]
\centering
\caption{Characteristic frequencies of spray wave, column flutter, column-breakup length and waves on column.}
\label{tab:freq}
\begin{tabular}{cccccc}
\hline
                 & \begin{tabular}[c]{@{}c@{}}Spray\\ wave\end{tabular} & \begin{tabular}[c]{@{}c@{}}Column\\ fluttering\end{tabular} & \multicolumn{2}{c}{Column length} & \begin{tabular}[c]{@{}c@{}}Waves on \\ column\end{tabular} \\ \hline
Frequency ($Hz$) & $\approx3100$                                        & $\approx3300$                                               & $f_1\approx1120$   & $f_2\approx16800$   & $\approx14830.5$                                           \\[0.1cm] 
Wavelength       & $\approx9D$                                                 & -                                                           & \multicolumn{2}{c}{-}             & $\approx D$                                                       \\[0.1cm] 
\hline
\end{tabular}
\end{table}





 






\section{Trajectory\label{sec:traj}}



The trajectory of a liquid jet in cross flow is a vital design parameter for the construction of a gas turbine combustor. In the present study, we quantify the trajectories of the jets for both the swirling and the non-swirling cases. \cite{wu1997breakup} proposed that the trajectory of a resultant spray can be given by a correlation expression,

\begin{equation}
r = a Q ^{0.5} x^b,
\label{equ:traj}
\end{equation}

where $r$ and $x$ are radial and axial distances normalized by the jet diameter, $a=0.55$ and $b=0.5$ are the coefficients. This correlation has been agreed upon by the subsequent experimental studies \citep{wu1997breakup,wu1998spray}. Though there were few revisions to this expression in terms of the minor modifications to the coefficients involved, the essential form has remained the same. Clearly, from the expression in Equation \ref{equ:traj}, the trajectory is dependent only on the liquid-to-gas momentum-flux ratio, $Q$, and not on the aerodynamic Weber number, $We_g$. Therefore, while $We_g$ dictates the transitions between the regimes of breakup, $Q$ influences the trajectory of the spray. Note here that, while the trajectories could be derived to represent either the centerline or the outer-windward boundary of the spray, in the present study we choose to use the latter. Although the trajectories obtained using the centerline or the outer-windward boundary of the spray are not equivalent, however, any one definition can be used to serve the purpose of characterizing the jet deformation provided that it is consistently used across all the calculations and in comparisons.

\begin{figure}[t!]
\centering
\includegraphics[width=0.65\textwidth]{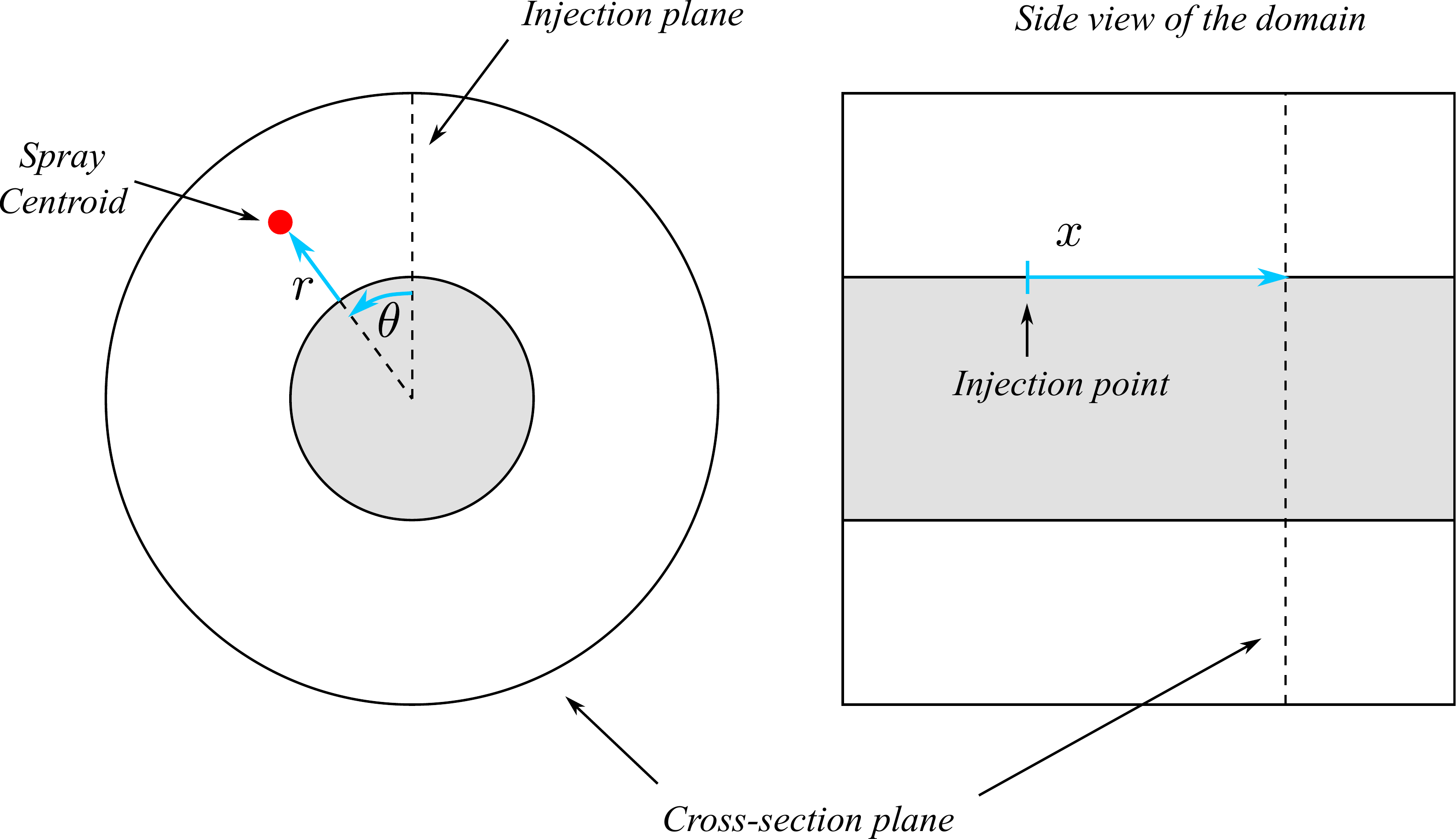}
\centering
\caption{Coordinates for trajectory calculation.}
\label{fig:coord}
\end{figure}

In the annular configuration, the spray traverses a three dimensional path which implies that a single expression is not sufficient to completely represent the trajectory. Therefore, the trajectory is expressed in terms of both the radial penetration and the angular deflection of the centroid of the drops at a streamwise location. Figure \ref{fig:coord} shows the schematic for the coordinates used for the calculation of the trajectory. Since, we use an annular geometry instead of the conventional rectangular cross-sections, the trajectory is represented in terms of the radial penetration with reference to the outer surface of the inner annular tube and is denoted by $r$. Angular deflection denoted by $\theta$ is measured with respect to the plane containing the point of liquid jet injection and the axis of the annular domain. 

The trajectory is then calculated by time-averaging the data to account for the crests and troughs formed on the outer windward boundary of the spray in the quasi-steady state. Experiments and previous numerical works rely on optical methods and image-processing methods to compute the trajectory, wherein the images obtained from the side view of the spray are averaged over time and then converted to gray scale to obtain the outer boundary of the spray and then a profile is fitted to the outer boundary to get the trajectory. However, this method is not general and cannot be used to obtain the trajectory of the spray in swirling-gas crossflows. So we choose to use the actual locations of the drops available from the simulation to obtain the trajectories. Instead of considering the location of a single droplet that has a maximum penetration in the radial direction as the boundary of the spray, we choose to consider the average of the location of a band of droplets that lie between 95\% and 99\% of the maximum penetration near the outer edge of spray as the boundary of the spray to avoid the presence of stray droplets influencing the trajectory calculation. 

\begin{figure}[t!]
\centering
\includegraphics[width=0.65\textwidth]{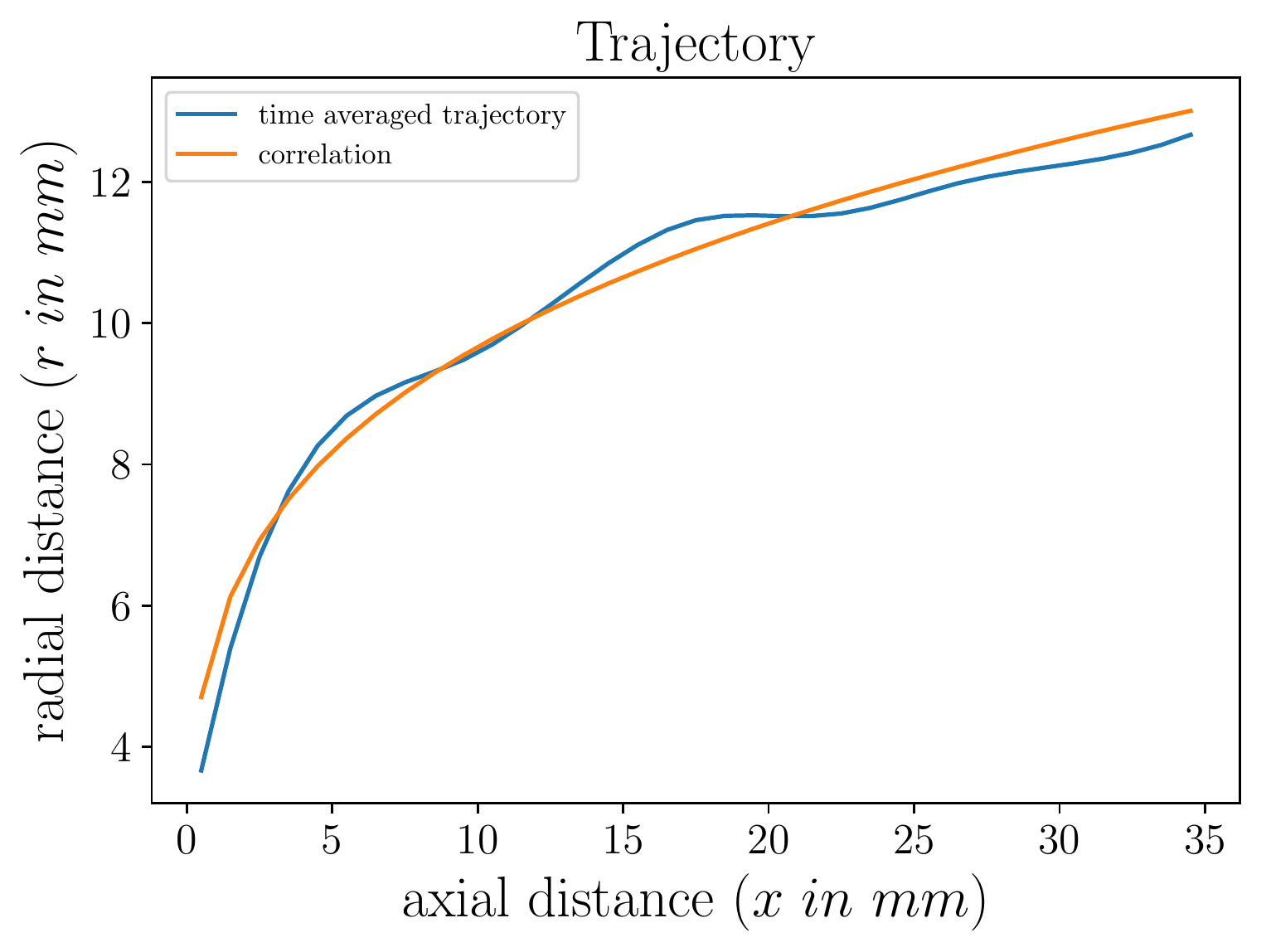}
\centering
\caption{Trajectory for q20sn0 case.}
\label{fig:traj}
\end{figure}

Figure \ref{fig:traj} shows the plot of time-averaged radial penetration for $q20sn0$ case in the present study, along with the best-fit curve, see Appendix A for the trajectory obtained from the simulations on a coarser grid. We use the same expression, the same dependence on $Q$ in Equation \ref{equ:traj}, for the radial penetration in all the cases and derived the correlation coefficients $a$ and $b$. Table \ref{tab:coeff} lists the derived correlation coefficients for all cases in the present study.

\begin{figure}[t!]
\centering
\includegraphics[width=0.65\textwidth]{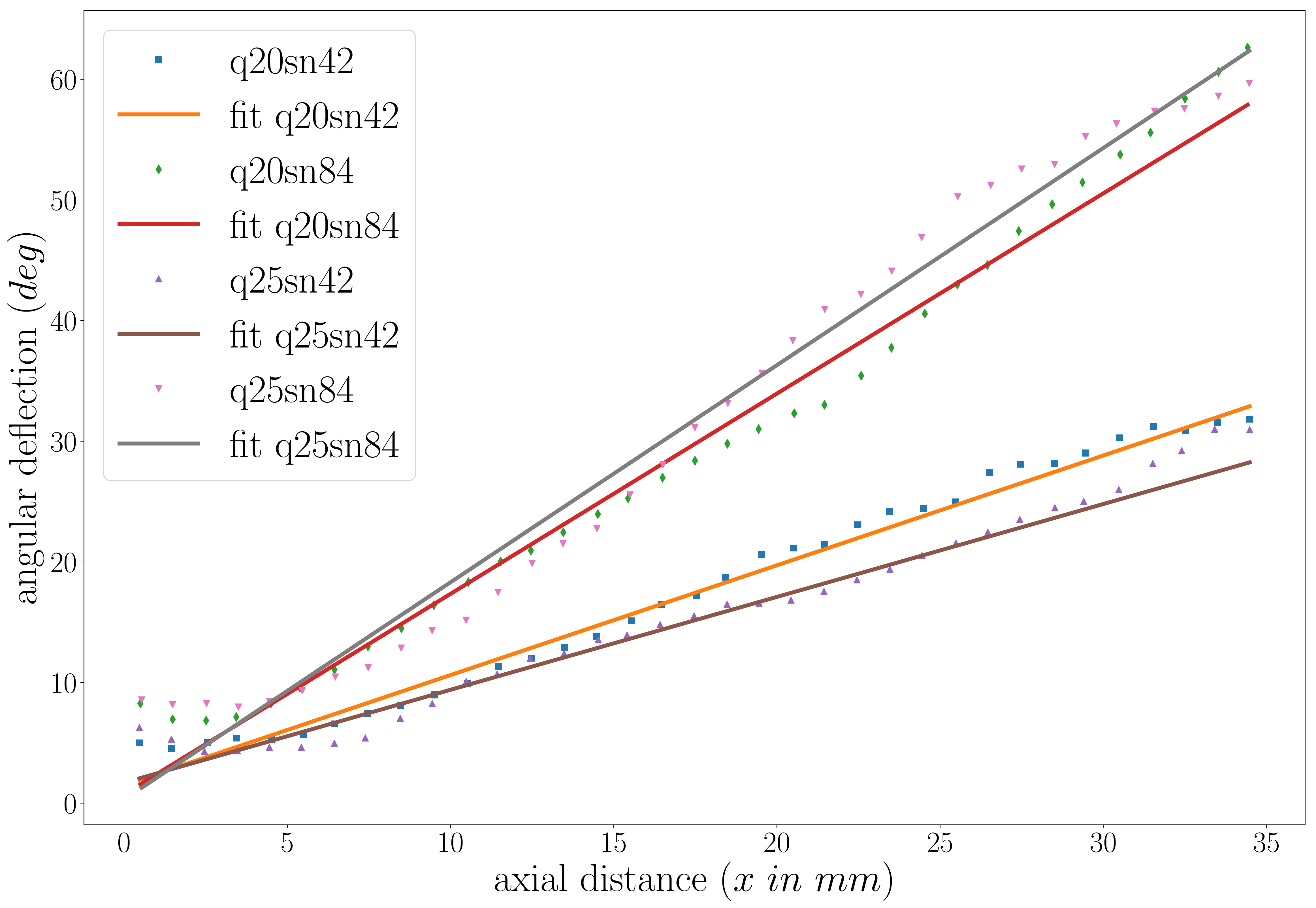}
\centering
\caption{Angular deflection.}
\label{fig:ad}
\end{figure}

Angular deflection of the centroid of the spray depends on the strength of the swirl component in the crossflow air. Hence, we chose to express the angular deflection as a general function of $SN$ and $x$, that is, $\theta=f(SN,x)$. Figure \ref{fig:ad} shows the plot of the time-averaged angular deflection of the centroid for all the cases in the present study, along with the best-fit line. Analyzing the simulation results, we find that $\theta$ varies linearly with $x$ and the slopes are proportional to the corresponding $SN$. Therefore, we propose the following expression for $\theta$ as,
\begin{equation}
\theta = k x + c,
\label{equ:deflect}
\end{equation}
where $k=2SN$ is the slope which was found empirically from the data and $c$ is the intercept. Table \ref{tab:coeff} lists the derived correlation coefficients $k$ and $c$ for all cases in the present study. The radial penetration together with the angular deflection fully describes the trajectory of the spray. Interestingly, the linear correlation with $\theta$ is also observed in experiments as discussed in \cite{prakash2017e}.

\begin{table}[H]
\centering
\caption{List of correlation coefficients for radial penetration and angular deflection of the centroid for all the cases.}
\label{tab:coeff}
\begin{tabular}{c|cccc|cccc}
\hline
 & \multicolumn{4}{c}{q20} & \multicolumn{4}{c}{q25} \\ 
SN    & a   & b   & k   & c  		& a & b & k & c			\\[0.1cm] \hline 
0     & 1.243 & 0.24 &  &  			& 1.376 & 0.22 &  &			\\[0.1cm] 
0.42  & 1.23 & 0.20 & 0.91 & 1.52 	& 1.32 & 0.21 & 0.77 & 1.7 		\\[0.1cm] 
0.84  & 1.21 & 0.20 & 1.66 & 0.75 	& 1.2356 & 0.21 & 1.8 & 0.31 		\\ \hline
\end{tabular}
\end{table}

\section{Drop size, velocity and shape-factor distributions\label{sec:distribution}}



\begin{figure}[t!]
\centering
\includegraphics[width=\textwidth]{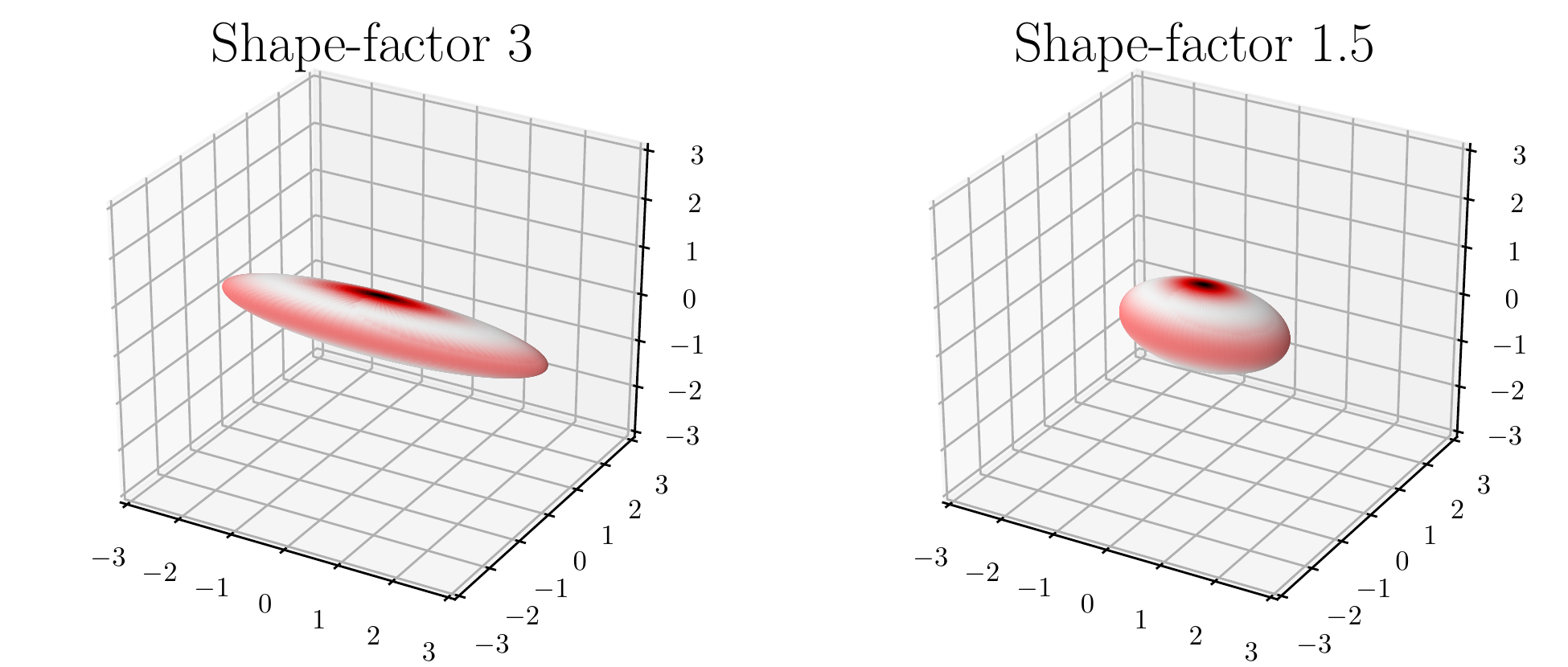}
\caption{Illustration of drop shapes for shape-factor 3 and 1.5. Shape-factor of 3 yields a prolate spheroid with length-to-breadth ratio of $\approx5.1$ and shape-factor of 1.5 yields a prolate spheroid with length-to-breadth ratio of $\approx1.85$.}
\label{fig:spheroid}
\end{figure}

In this section, we present the study of drop size, velocity and shape-factor distribution of the resultant spray. The larger drops and ligaments formed during the primary atomization undergo further breakup into smaller droplets completing the breakup through the secondary atomization.  The variation of drop sizes and their velocities along the stream-wise direction are important parameters to be studied. 
The shape factor is another important parameter used to characterize the shape of the droplets formed and is defined as the ratio of largest radius (from centroid) to the mean radius of a liquid drop. This is an alternative measure to the conventionally used parameter, sphericity defined as the area of the sphere with the equivalent diameter to the actual area of the sphere. The choice of the sphericity parameter used in this study to represent the shape of a drop yields a wider range of numbers that helps in effectively classifying the ligaments and near-spherical droplets. Figure \ref{fig:spheroid} illustrates the representative shapes of the drops for shape-factor 1.5 and 3.

\begin{figure}[t!]
\centering
\includegraphics[width=\textwidth]{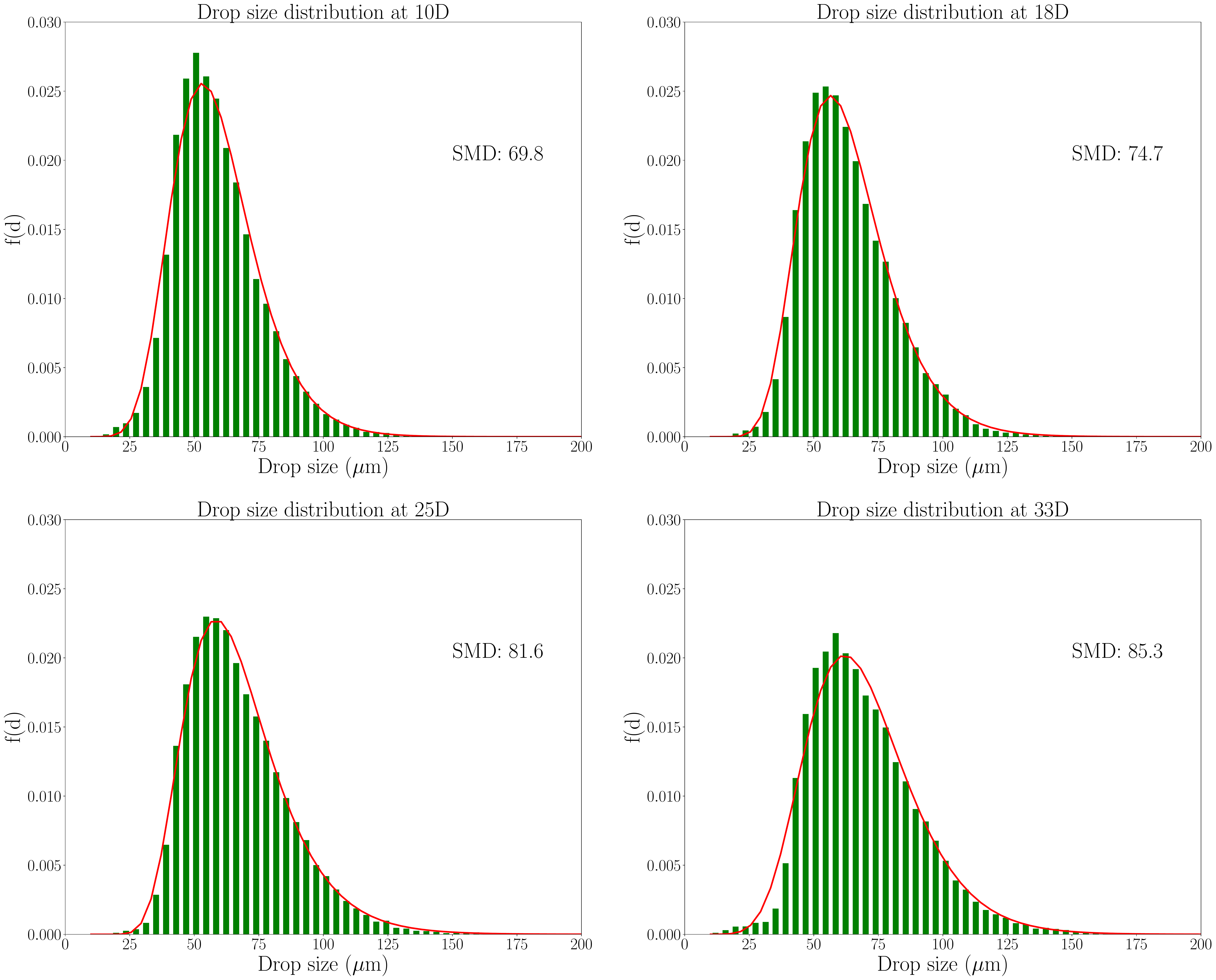}
\centering
\caption{Probability density function for drop size $f(d)$ for the $q20sn0$ case at downstream locations of 10D, 18D, 25D and 33D. Solid line represents the log-normal fit.}
\label{fig:size_q20sn0}
\end{figure}
\begin{figure}[t!]
\centering
\includegraphics[width=0.6\textwidth]{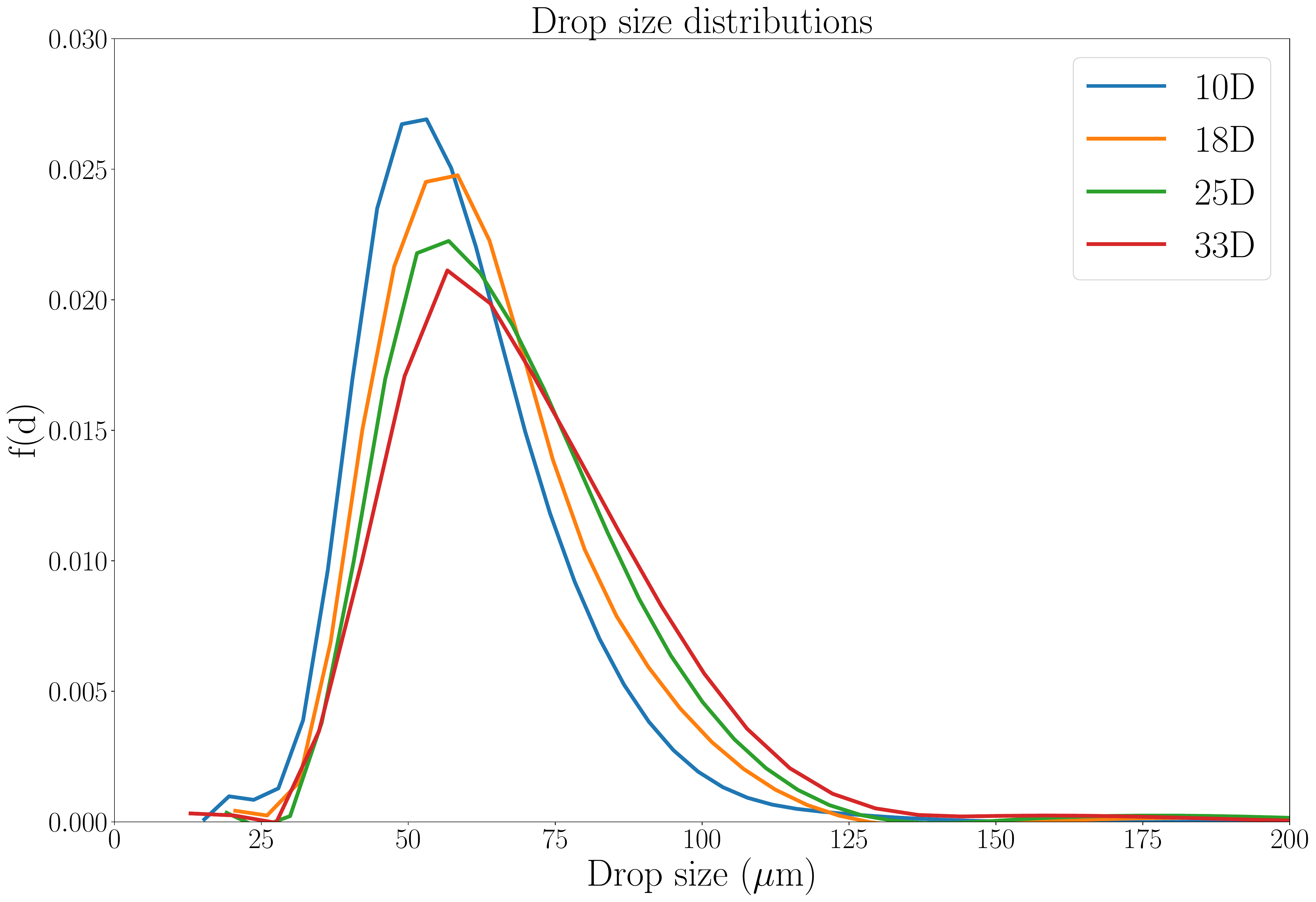}
\centering
\caption{Probability density function for drop size $f(d)$ for the $q20sn0$ case at downstream locations of 10D, 18D, 25D and 33D.}
\label{fig:size_q20sn0_all}
\end{figure}

Figure \ref{fig:size_q20sn0} shows the probability density function of drop size $f(d)$ for the $q20sn0$ case at four downstream locations $10D$, $18D$, $25D$ and $33D$. We considered all the droplets irrespective of the value of shape factor in the drop-size analysis, though drops with shape-factor values higher than 3 can be mostly considered as ligaments and can be excluded from the analysis as was done in \citet{prakash2016} - see Appendix B for the sensitivity of results on the choice of threshold based on shape-factor values. We observed that the drop sizes obtained follow the log-normal distribution consistent with the previous experimental observations \citep{adebayo2015}. This observation is found to be consistent at all the locations for both the swirling and non-swirling cases. Tables \ref{tab:smd_q20} and \ref{tab:smd_q25} lists the summary of the study of variation of drop-sizes along the streamwise direction for all the cases studied here. It is clear that the Sauter-mean diameter (SMD) of the drops monotonically increases in magnitude as the drops move downstream towards the exit of the domain. Figure \ref{fig:size_q20sn0_all} also shows that the probability density function for drop size, $f(d)$, shifts to the right. We attribute this behavior to the coalescence of drops occurring during the downstream motion of the droplets. 

\begin{table}[H]
\centering
\caption{SMD values for q20.}
\label{tab:smd_q20}
\begin{tabular}{c|cccc}
\hline
SN   & 10D   & 18D   & 25D   & 33D   \\ \hline
0    & 69.81 & 74.71 & 81.59 & 85.25 \\[0.1cm] 
0.42 & 73.13 & 78.87 & 83.47 & 86.47 \\[0.1cm] 
0.84 & 75.24 & 79.42 & 89.11 & 90.83 \\ \hline
\end{tabular}
\end{table}

\begin{table}[H]
\centering
\caption{SMD values for q25.}
\label{tab:smd_q25}
\begin{tabular}{c|cccc}
\hline
SN   & 10D   & 18D   & 25D   & 33D   \\ \hline
0    & 74.12 & 79.32 & 81.16 & 86.19 \\[0.1cm] 
0.42 & 74.06 & 79.97 & 82.11 & 87.65 \\[0.1cm] 
0.84 & 75.06 & 83.42 & 87.16 & 89.10 \\ \hline
\end{tabular}
\end{table}

\begin{figure}[H]
\centering
\includegraphics[width=0.8\textwidth]{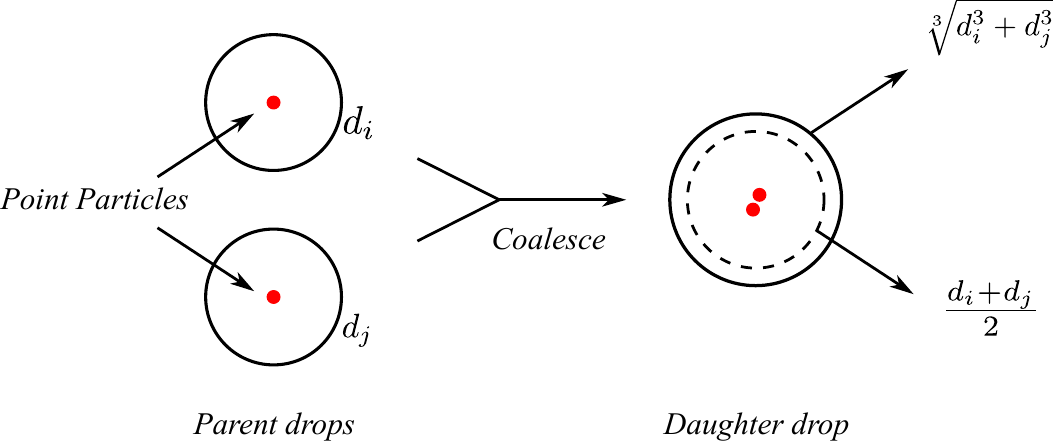}
\centering
\caption{Schematic showing the setup to verify coalescence.}
\label{fig:coalesce}
\end{figure}

\begin{figure}
    \centering
    \includegraphics[width=0.7\textwidth]{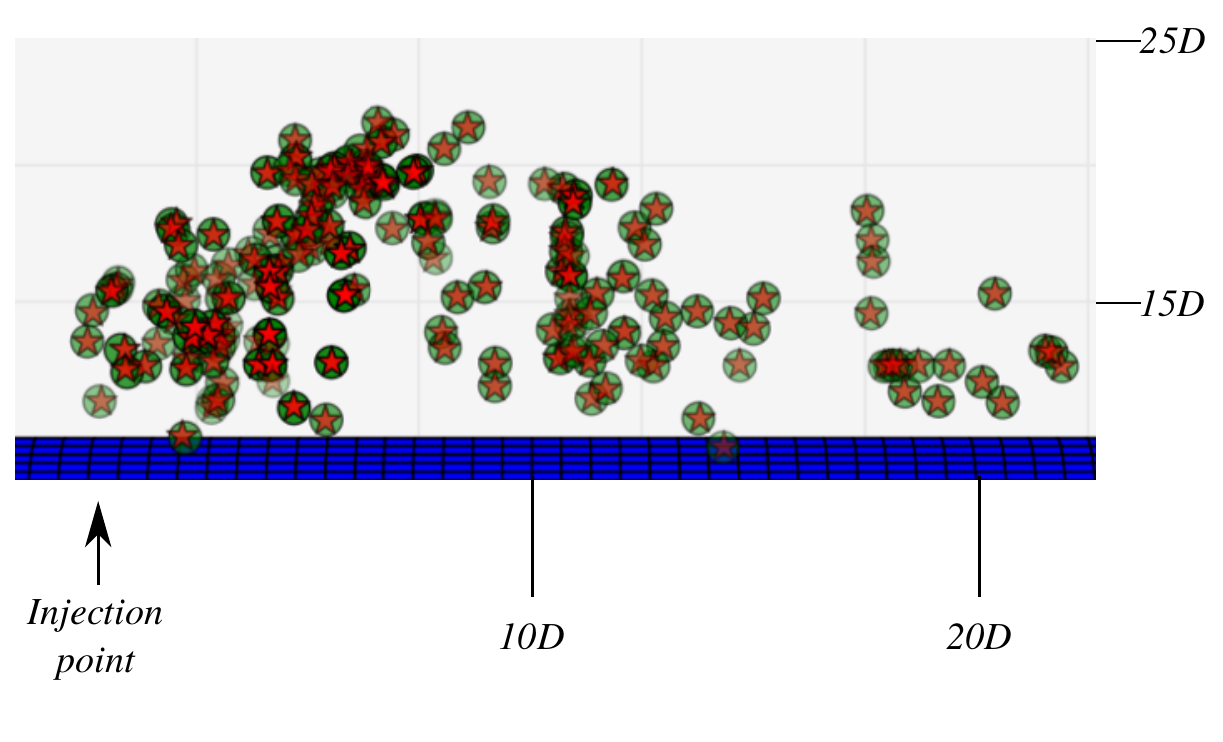}
    \caption{Location of parent drops (red stars and green circles) that undergo coalescence and result in a single drop between $t_i=750\mu s$ and $t_l=775\mu s$.}
    \label{fig:coalesce_example}
\end{figure}

To further verify our claim, we adopt the point-particle model in Gerris solver \citep{tomar2010multiscale} and modify it to detect the coalescence of droplets in the domain. At an arbitrary chosen time $t_i$ in the quasi-steady state of the spray, we add a passive-point particles at the centroid of each of the drops throughout the domain and tag all the particles with an ID number and the diameter $d$ of the drops that they belong to. Particles were then advected with the background velocity field. At any later time $t_l$, we define that the coalescence is said to have occurred between the two drops of diameters $d_i$ and $d_j$ if the two associated point particles are separated by a distance of $l$ that is bound by the mean diameter given as,
\begin{equation}
l<\frac{(d_i + d_j)}{2}. 
\label{equ:dist}
\end{equation}
This is schematically shown in Figure \ref{fig:coalesce}. The bound used in the Equation \ref{equ:dist} is more stricter than the bound that could be picked based on the sum of the volumes of the two parent drops i.e $\sqrt[3]{d_i^3 + d_j^3}$ as also schematically represented in the Figure \ref{fig:coalesce}. Further to verify that this condition indeed represents a coalescence of two drops, all pairs of particles that satisfy the condition in the Equation \ref{equ:dist} at a time $t_l$ are tested to satisfy the condition at another time $t=t_l+\delta t$ assuming that there is no breakup for a small time $\delta t$, verifying that the particles stick together once the parent drops coalesce to form a daughter drop. For example, at an arbitrary time instance of $t_i\approx750\mu s$, VOF resolved drops were converted into point particles and at a later time $t_l=775\mu s$, coalescence was tested. A total of 185 coalescence instances were found in a time interval of $25\mu s$ and the location of these events are shown in Figure \ref{fig:coalesce_example}, proving the existence of coalescence. We performed this exercise to confirm the presence of coalescence and to show that the increase in SMD seen in the Tables \ref{tab:smd_q20} and \ref{tab:smd_q25} could be due to the coalescence of drops as they move downstream in the domain. However, in reality breakup and coalescence dynamics of drops are much more complicated and tightly coupled. Higher amount of coalescence could result in larger drops and larger drops could in turn lead to a higher amount of breakup and hence the resultant drop sizes are in a way dependent on both breakup and coalescence dynamics of drops and the complex interplay between them. 

\begin{figure}[H]
\centering
\includegraphics[width=\textwidth]{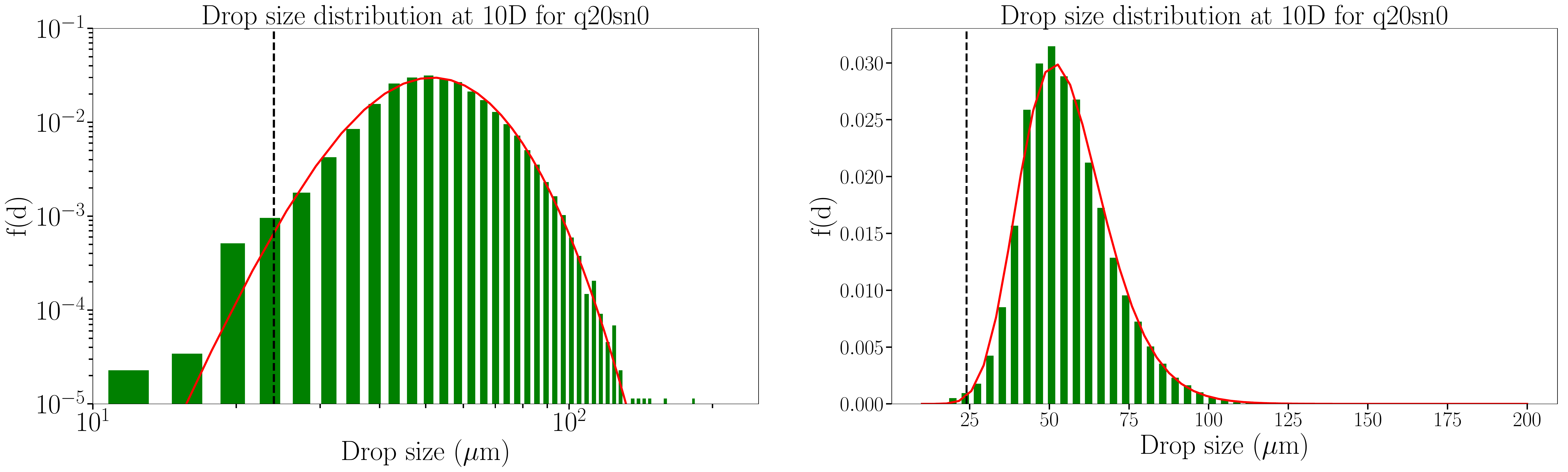}
\centering
\caption{Probability density function for drop size $f(d)$ in logarithmic scale (left) and linear scale (right), showing the well-resolved left arm of the distribution. Solid line is the log-normal fit. Dashed line represents the grid size.}
\label{fig:grid}
\end{figure}

The smooth variation of the left arm of the drop-size distribution curve suggests that the employed grid-size resolution sufficiently captures the dominant secondary breakup of droplets. Figure \ref{fig:grid} shows the plot of the drop-size distribution for q20, sn0 case in logarithmic space and linear space. Vertical-dashed line represents the grid size. Clearly, we capture most of the drops on the left arm of the distributions and we can observe from the plot in log space that all the resolved drops follow the log-normal distribution very well, illustrating the high fidelity of our simulation results. However, one has to be aware of the shortcomings of the VOF methods. VOF methods generate a large amount of floatsam and jetsam that are of size comparable to or smaller than the grid size and these spurious drops have the potential to alter the drop size distribution. In a recent study, \cite{Li2016SprayDrops} discussed this issue in detail. Hence, the quantitative drop size distributions obtained from any resolved numerical simulation does not result in the complete spectrum of droplets. However, these results are still very valuable in understanding the essential mechanisms of jet breakup and spread of the droplets in the crossflow.    

\begin{figure}
    \centering
    \includegraphics[width=0.5\textwidth]{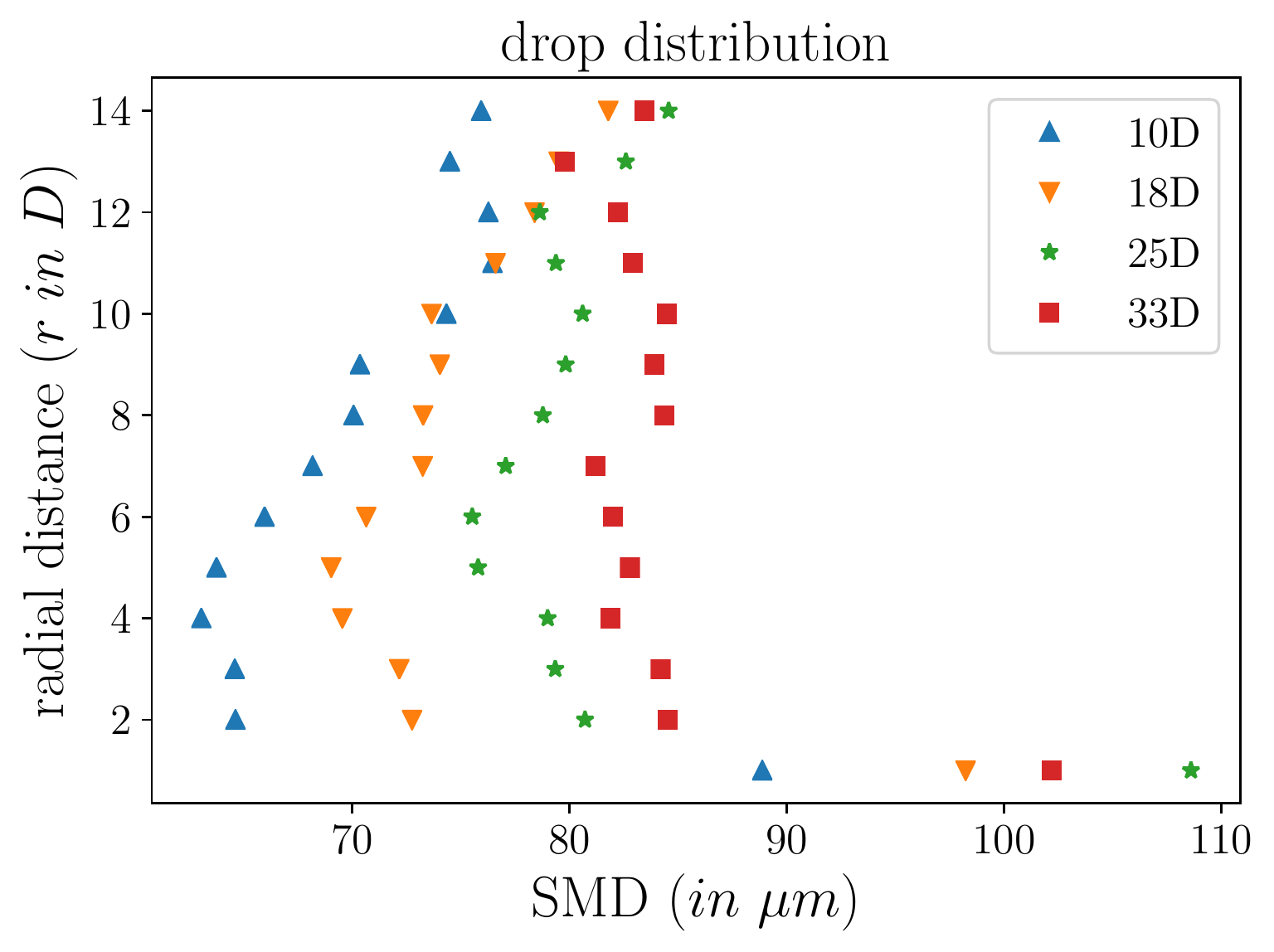}
    \caption{Time averaged drop size values at different radial locations and at axial locations of $10D$, $18D$, $25D$ and $33D$ from the jet, for the $q20sn0$ case.}
    \label{fig:radial}
\end{figure}

We also studied the variation in the drop-size distribution along the radial direction \citep{prakash2016} and the values at the downstream locations of $10D$, $18D$, $25D$ and $33D$ from the jet are shown in the Figure \ref{fig:radial} for $q20sn0$ case. The drop-size distribution shifts to the right from a downstream location of $10D$ to $33D$. At $10D$, drop size values decrease radially outwards, whereas at $33D$ drop sizes are fairly uniform. At locations very close to the inner cylinder, large SMD values can be seen at all locations which is due to the stripping of drops from the surface of the cylinder which leads to the formation of elongated ligaments (see Figure \ref{fig:renderside}) resulting in large SMD values. A similar qualitative behavior of drop-size distribution along the radial direction is seen across all cases. Since, this increase in drop sizes downstream is seen in all cases, we believe that the annular geometry leads to a complex flow that promotes coalescence. Later in the section, we show evidence for the drop coalescence events. Also, the effect is enhanced for higher Swirl numbers (see Tables \ref{tab:smd_q20} and \ref{tab:smd_q25} also shows that the SMD of drops increases with increase in SN) due to the following reason. This could be due to the coalescence of larger drops (radially traveling towards the periphery of the jet due to centrifugal forces) with other drops. However, a more rigorous analysis is required to prove this observed behavior which will be undertaken in near future. We note here that, these coalescence events do not occur for uniform flow in a rectangular cross-section channel.

\begin{figure}[t!]
\centering
\includegraphics[width=\textwidth]{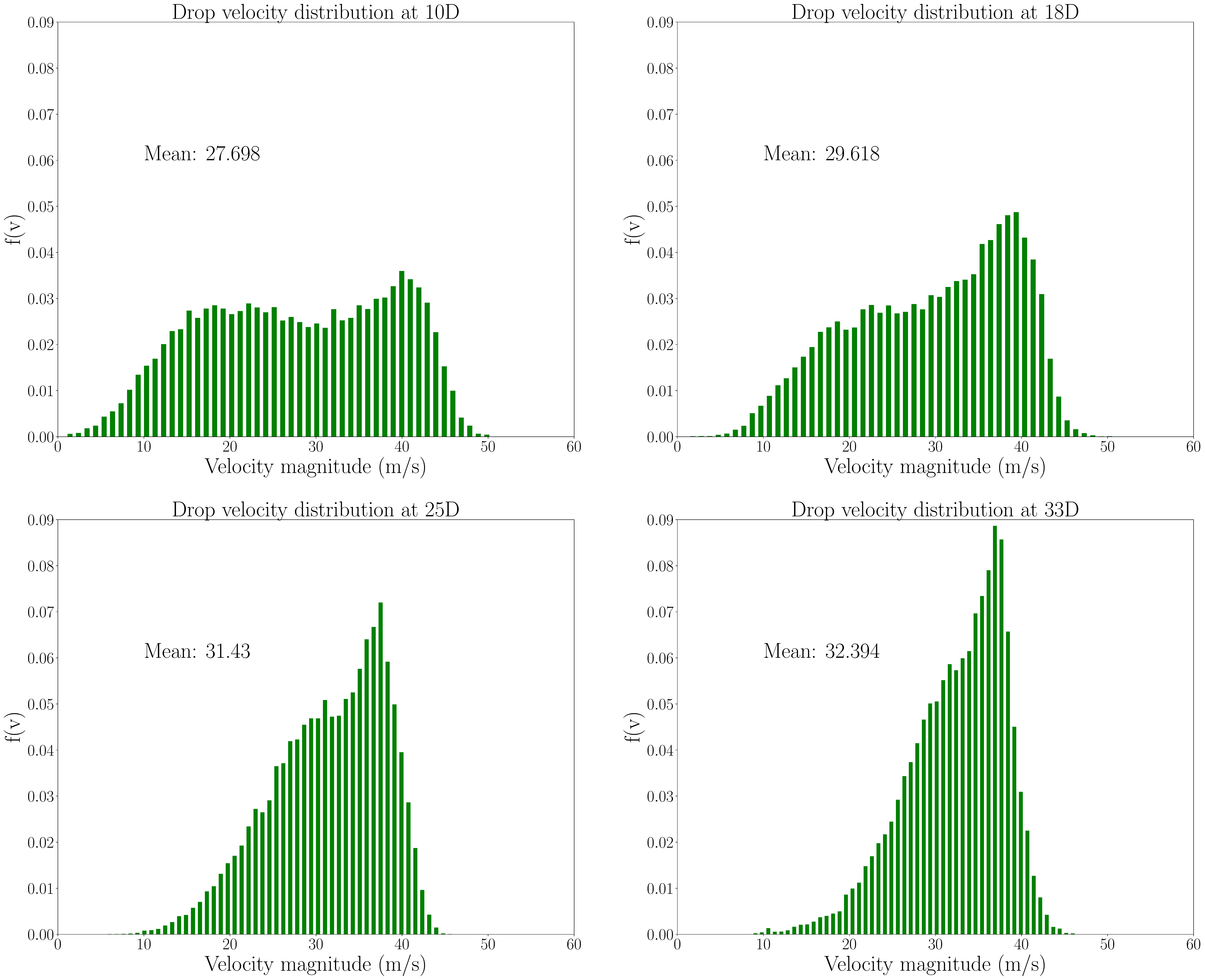}
\centering
\caption{Probability density function for drop velocity $f(v)$ for the $q20sn0$ case at downstream locations of 10D, 18D, 25D and 33D.}
\label{fig:vel_q20sn0}
\end{figure}

\begin{figure}[t!]
\centering
\includegraphics[width=0.6\textwidth]{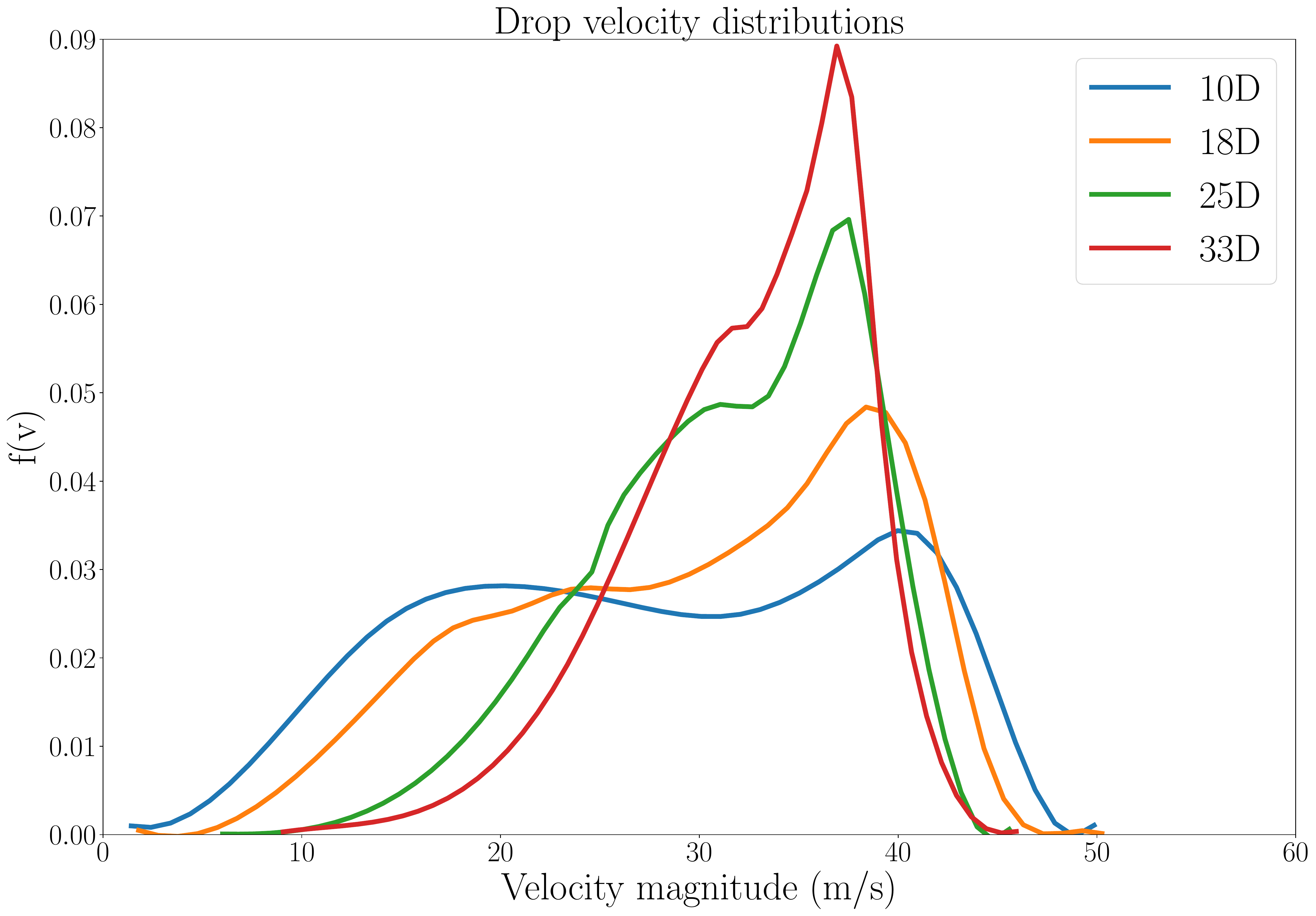}
\centering
\caption{Probability density function for drop velocity $f(v)$ for the $q20sn0$ case at downstream locations of 10D, 18D, 25D and 33D.}
\label{fig:vel_q20sn0_all}
\end{figure}

\begin{figure}[t!]
\centering
\includegraphics[width=\textwidth]{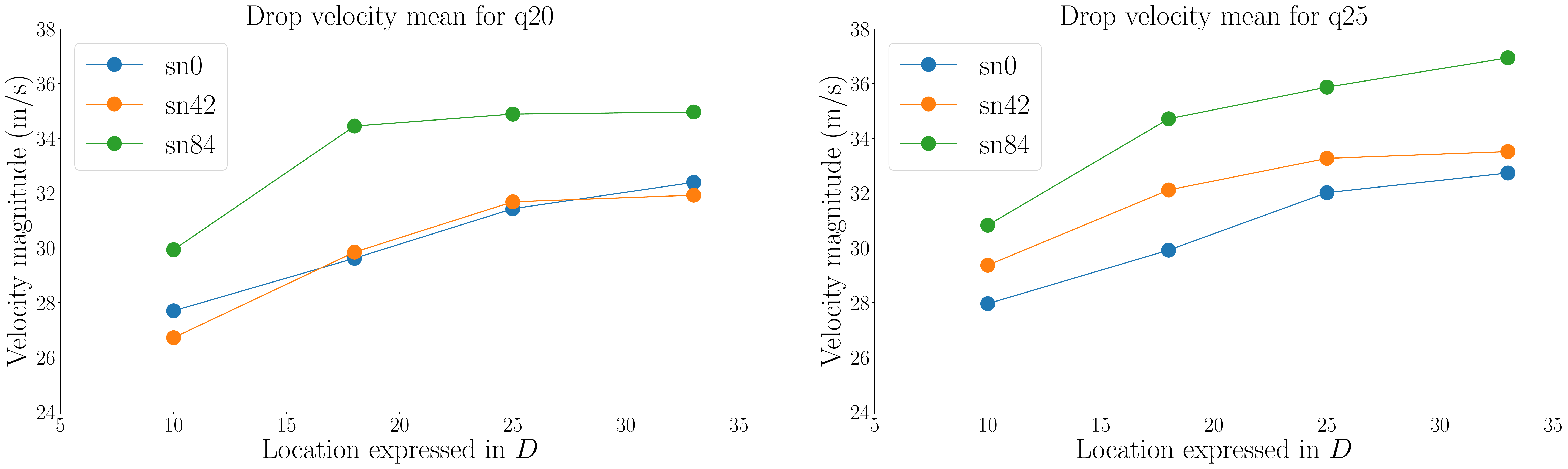}
\centering
\caption{Drop velocity mean for all cases at downstream locations of 10D, 18D, 25D and 33D.}
\label{fig:vel_mean}
\end{figure}

Velocities are also measured at the four downstream locations of $10D$, $18D$, $25D$ and $33D$ and the probability density function of the magnitudes ($||u||_2$) are plotted in the Figure \ref{fig:vel_q20sn0} for the $q20sn0$ case. We observe from the mean values in the plot that there is a monotonic increase in the velocity magnitude of the droplets as they move downstream along the flow. Interestingly, a bimodal-velocity distribution is observed at $10D$, which could be due to the lower axial velocity of the drops in the wake of the jet. However, the drops with low velocity are seen to accelerate as they move downstream to reach the free-stream velocity and the drops with higher velocity are seen to slow down due to drag, attaining a unimodal-velocity distribution at $33D$. Figure \ref{fig:vel_q20sn0_all} clearly shows this shift in the behavior of drops from $10D$ to $33D$. Figure \ref{fig:vel_mean} shows the mean of velocity magnitude for all cases in the present study. A similar behavior of increasing mean velocity values along the downstream can be seen for all the six cases. With an increase in $SN$, we see that the mean velocity is increasing for both q20 and q25 due to the higher total velocity experienced by the drops with the additional swirl components. Interestingly, we observe that the velocities for $q20sn42$ case are similar in magnitude compared to that for $q20sn0$ case especially since for $q25sn42$ case an increase in velocity is observed over $q25sn0$ as expected. It is not clear at this stage why this occurs and a more detailed analysis is required to explain this.

\begin{figure}[t!]
\centering
\includegraphics[width=\textwidth]{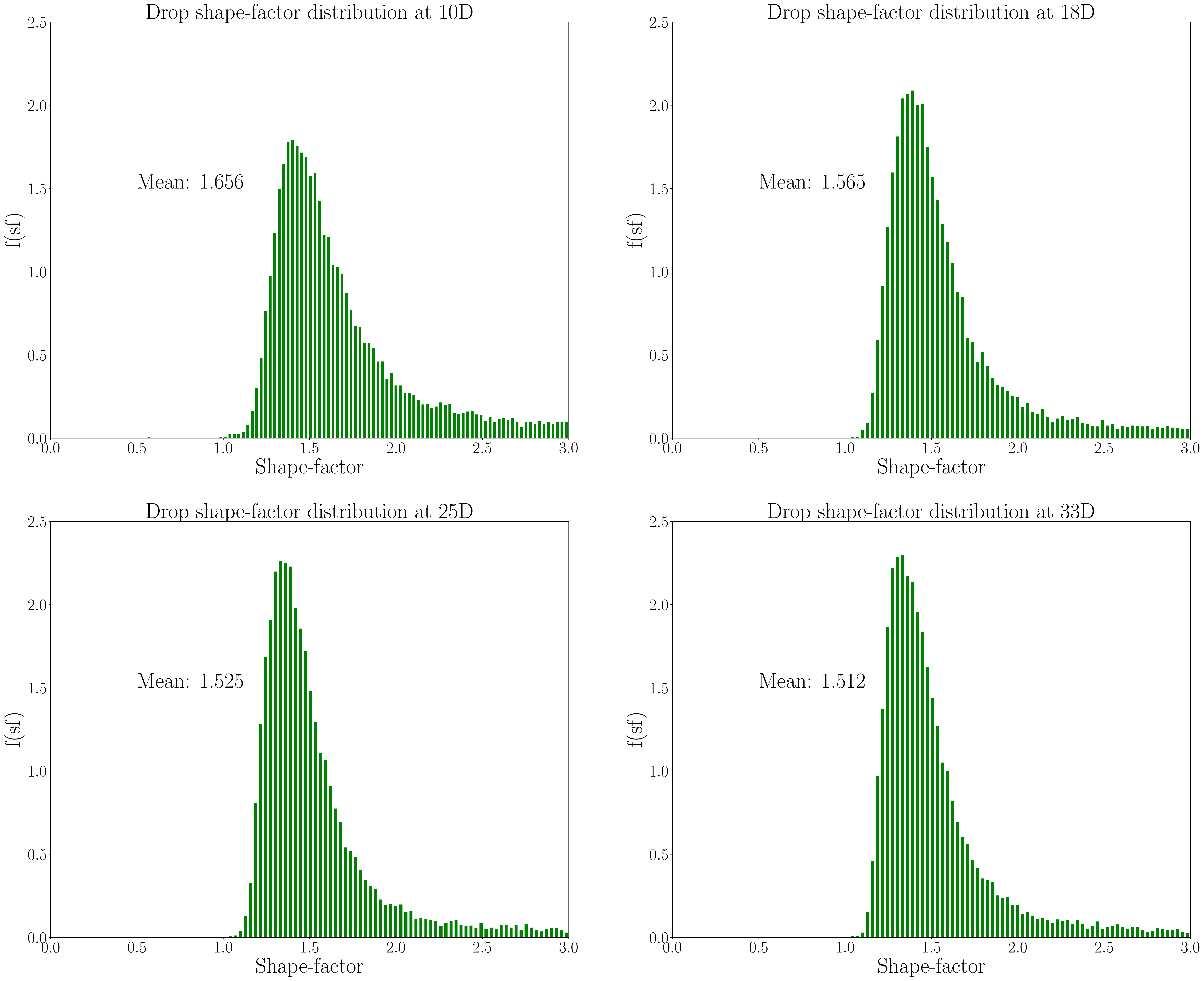}
\centering
\caption{Probability density function for drop shape-factor $f(sf)$ for the $q20sn0$ case at downstream locations of 10D, 18D, 25D and 33D.}
\label{fig:shape_q20sn0}
\end{figure}

\begin{figure}[t!]
\centering
\includegraphics[width=0.6\textwidth]{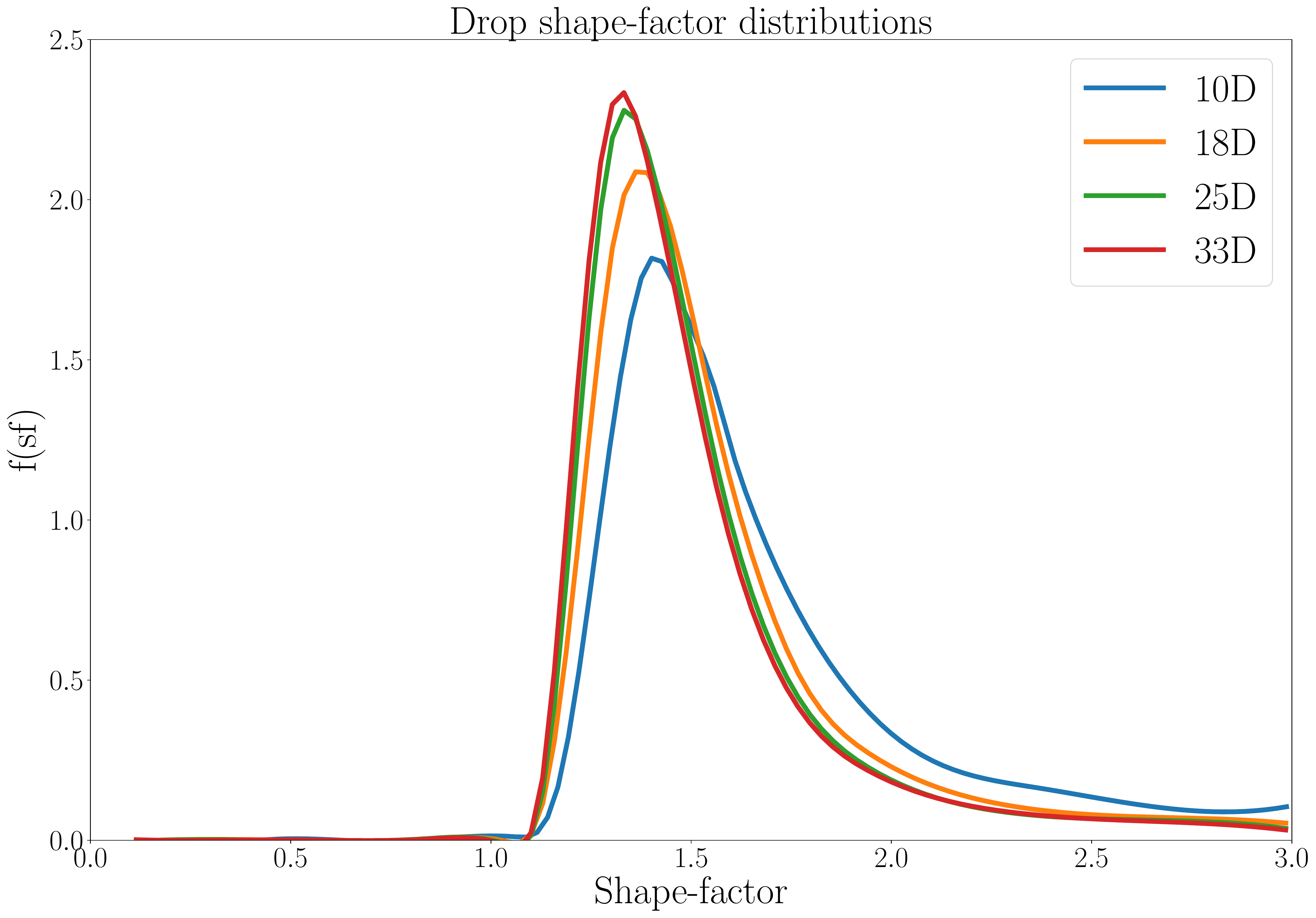}
\centering
\caption{Probability density function for drop shape-factor $f(sf)$ for the $q20sn0$ case at downstream locations of 10D, 18D, 25D and 33D.}
\label{fig:shape_q20sn0_all}
\end{figure}

\begin{figure}[t!]
\centering
\includegraphics[width=\textwidth]{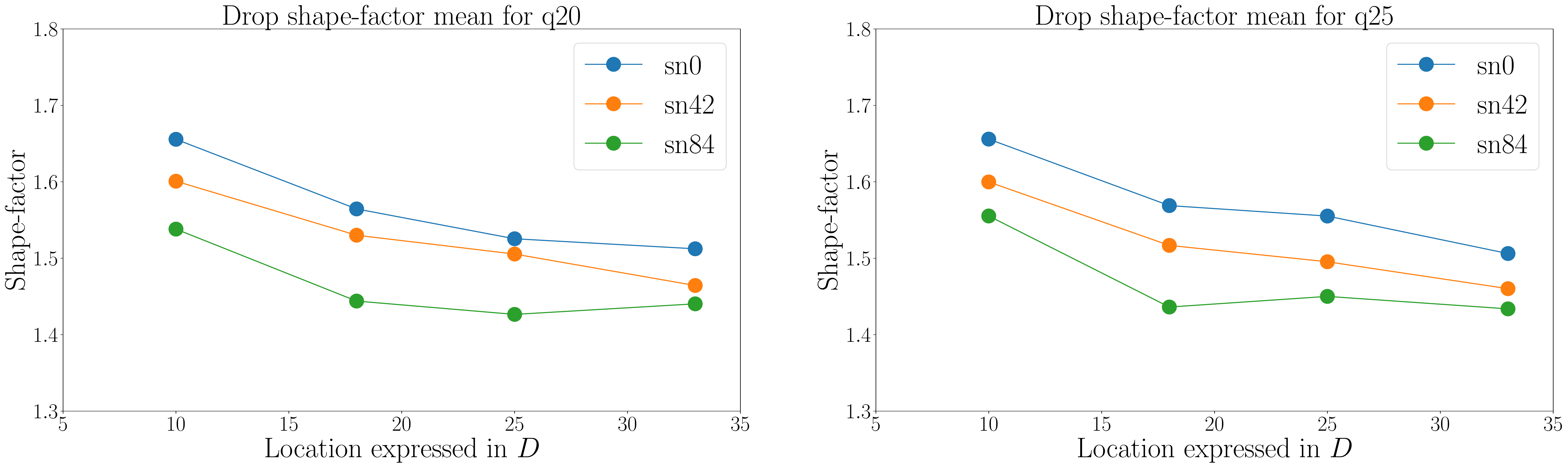}
\centering
\caption{Drop shape-factor mean for all cases at downstream locations of 10D, 18D, 25D and 33D.}
\label{fig:shape_mean}
\end{figure}

Shape factors are measured at four downstream locations of $10D$, $18D$, $25D$ and $33D$ and the values are plotted in Figure \ref{fig:shape_q20sn0} for the case of q20, sn0. We clearly observe from the mean values in the plot that there is a monotonic decrease in the shape factor of the droplets as they move downstream along the flow, which implies that the drops become more spherical due the action of surface tension force. Figure \ref{fig:shape_q20sn0_all} clearly shows the left shift in the probability density function of drop shape-factor from $10D$ to $33D$. Figure \ref{fig:shape_mean} shows the mean of shape-factor values for all the cases in the present study. A similar behavior of decreasing mean shape-factor values along the downstream can be seen for all the six cases. With an increase in $SN$, we can also see that the mean shape-factor is decreasing for both q20 and q25, which is essentially due to the longer distances traveled by the drops to reach the same axial location for higher values of $SN$.

\section{Conclusion\label{sec:conclusion}}

Numerical simulations of the breakup of a liquid-jet in a swirling crossflow were carried out at a density ratio of $180:1$, with swirl numbers of $0,0.42$ and $0.84$ that are equivalent to swirler vane angles of $0, 30$ and $45\ deg$. Both primary and secondary breakup have been analyzed to study the breakup modes. Drop size is observed to increase with the increase in swirl. The three-dimensional spray trajectory is expressed in terms of the radial penetration and the angular deflection, for which the correlations were obtained from the numerical simulations. The angular deflection is observed to vary linearly with the swirl number which agrees well with the experimental observations \citep{prakash2017e}. We note that the angular deflection is smaller than that of the swirling ambient flow. The penetration of the jet, in the radial direction, is observed to decrease with an increase in the swirl.

The predicted drop sizes are found to follow a log-normal distribution in agreement with the previous experimental findings. Sauter mean diameter of the droplet are observed to increase along the axial distance. We show that this behaviour is due to the coalescence of smaller droplets to form larger droplets. These observations clearly reflect the complex effects of the swirling cross flow in annular geometries.

\section*{Appendix A: Trajectory computed on different grids}

\begin{figure}
    \centering
    \includegraphics[width=0.75\textwidth]{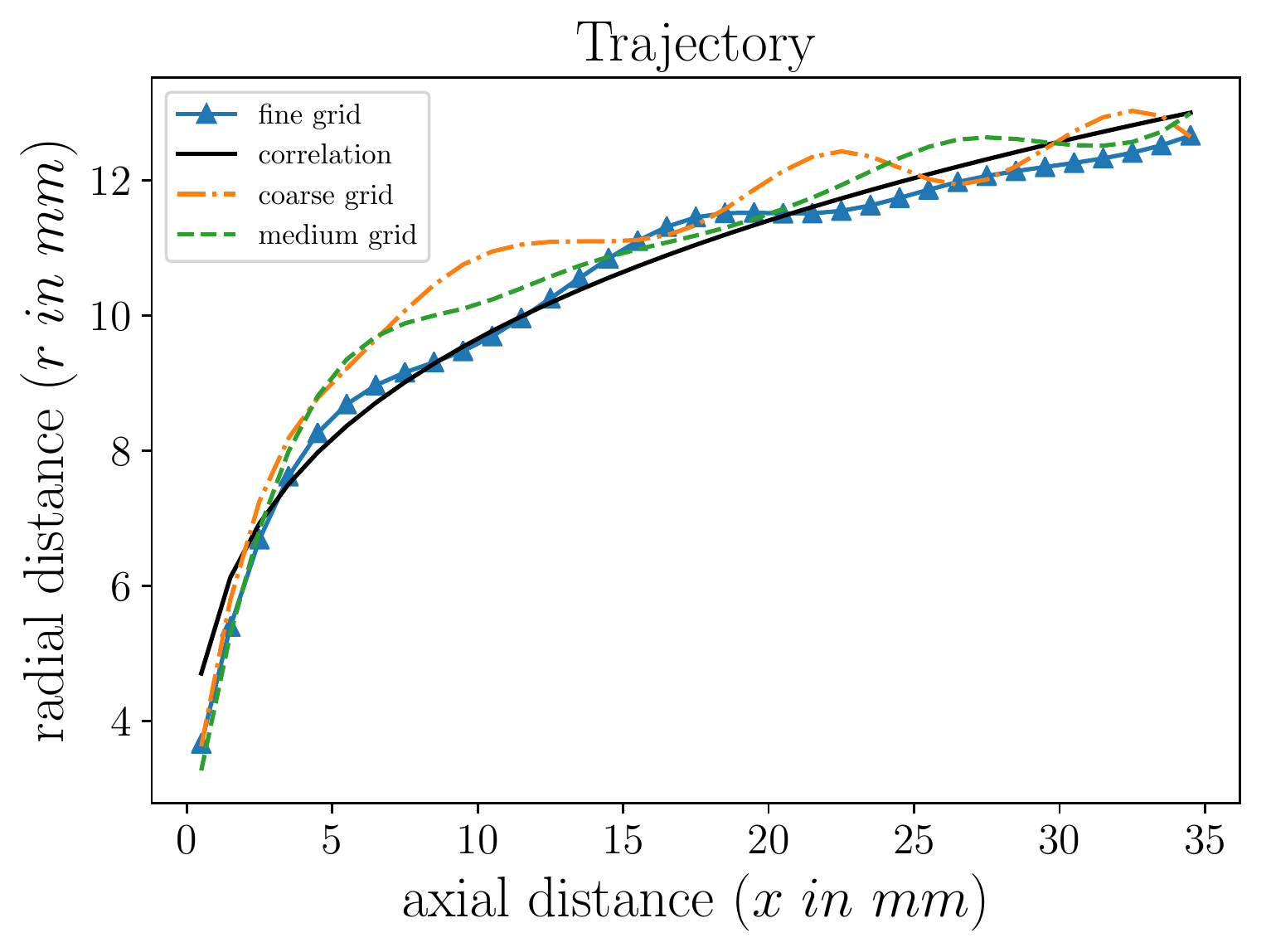}
    \caption{Trajectory for q20sn0 case on three different grids. }
    \label{fig:converge}
\end{figure}

In Section \ref{sec:traj}, we presented the time-averaged radial penetration for the $q20sn0$ case on our most refined grid, where the maximum refinement of 41 cells per diameter of jet is used in the regions close to the interface to resolve the intricate flow structures that are formed on the jet surface. However, it is in general a good practice to establish the invariance of results with respect to the refinement of the grid. Hence in this section, we present the trajectory of the jet in Figure \ref{fig:converge} on two coarser grids compared to the one used everywhere else in the paper. 
We call the ``fine grid" as the one that was used for all the results presented in the paper. The ``medium grid" is a grid with a refinement of 20 cells per diameter of the jet in the regions close to the interface which is 1 level coarser compared to the fine grid at the interface, however the surrounding flow that is refined based on the local vorticity field is maintained at the same level of refinement as that of the fine grid. The ``coarse grid" is a grid with a refinement of 20 cells per diameter of the jet in the regions close to interface and the surrounding flow that is refined based on the local vorticity field is also a level coarser compared to the fine grid. The agreement between the trajectories from different grids is fairly reasonable and can be considered as grid independent.

\section*{Appendix B: Sensitivity of the drop distributions on the shape-factor threshold value}

In Section \ref{sec:distribution}, we presented the results of drop size, velocity and shape-factor distributions by including all the drops. However, drops with a shape-factor value of roughly 3 and higher can be considered as ligaments and can be excluded from the analysis as was done in \citet{prakash2016}. However, it is important to quantify the sensitivity of the results on the threshold value of shape factor to determine the role played by these elongated droplets (or ligaments). 

\begin{figure}
    \centering
    \includegraphics[width=0.49\textwidth]{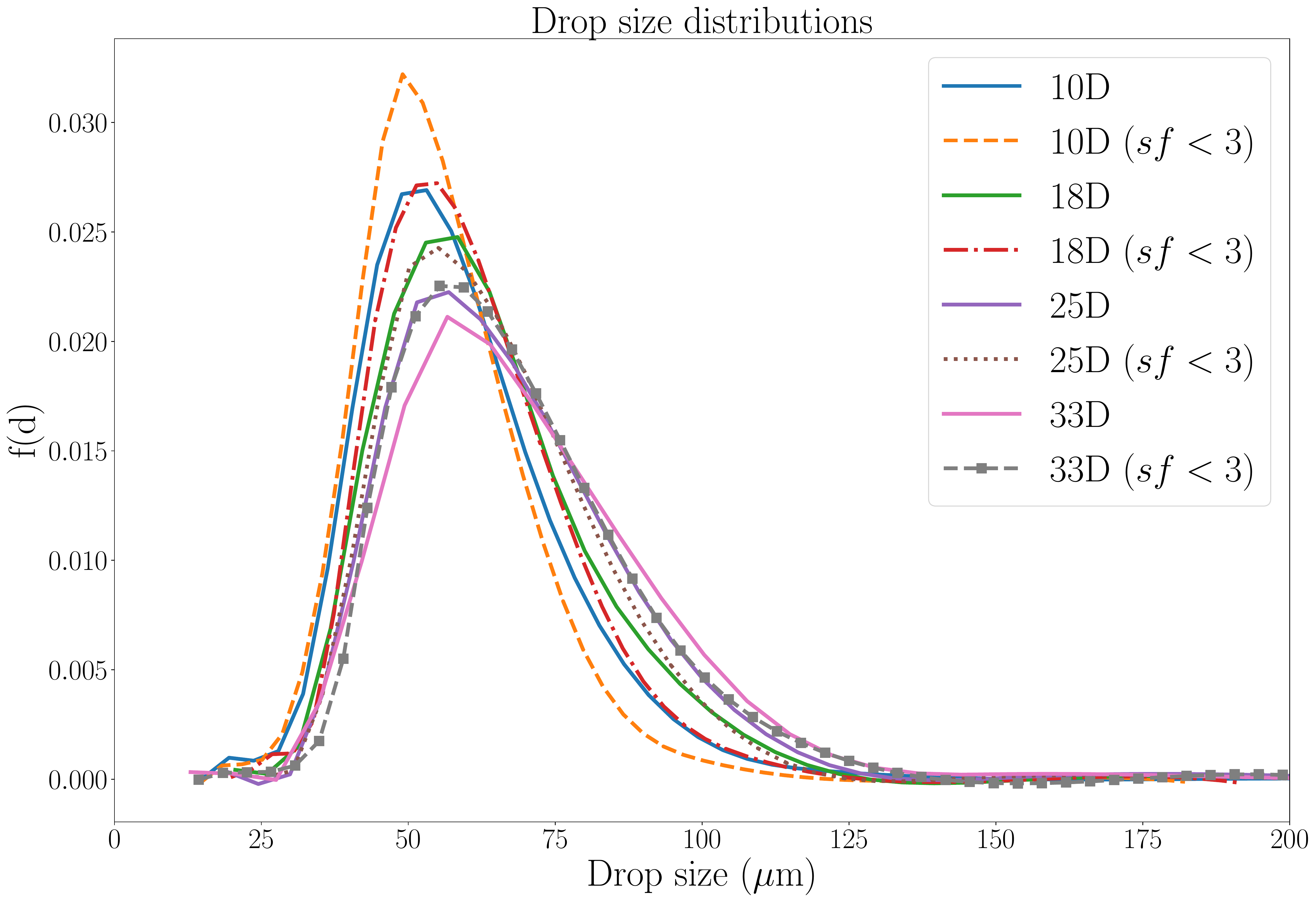}
    \includegraphics[width=0.49\textwidth]{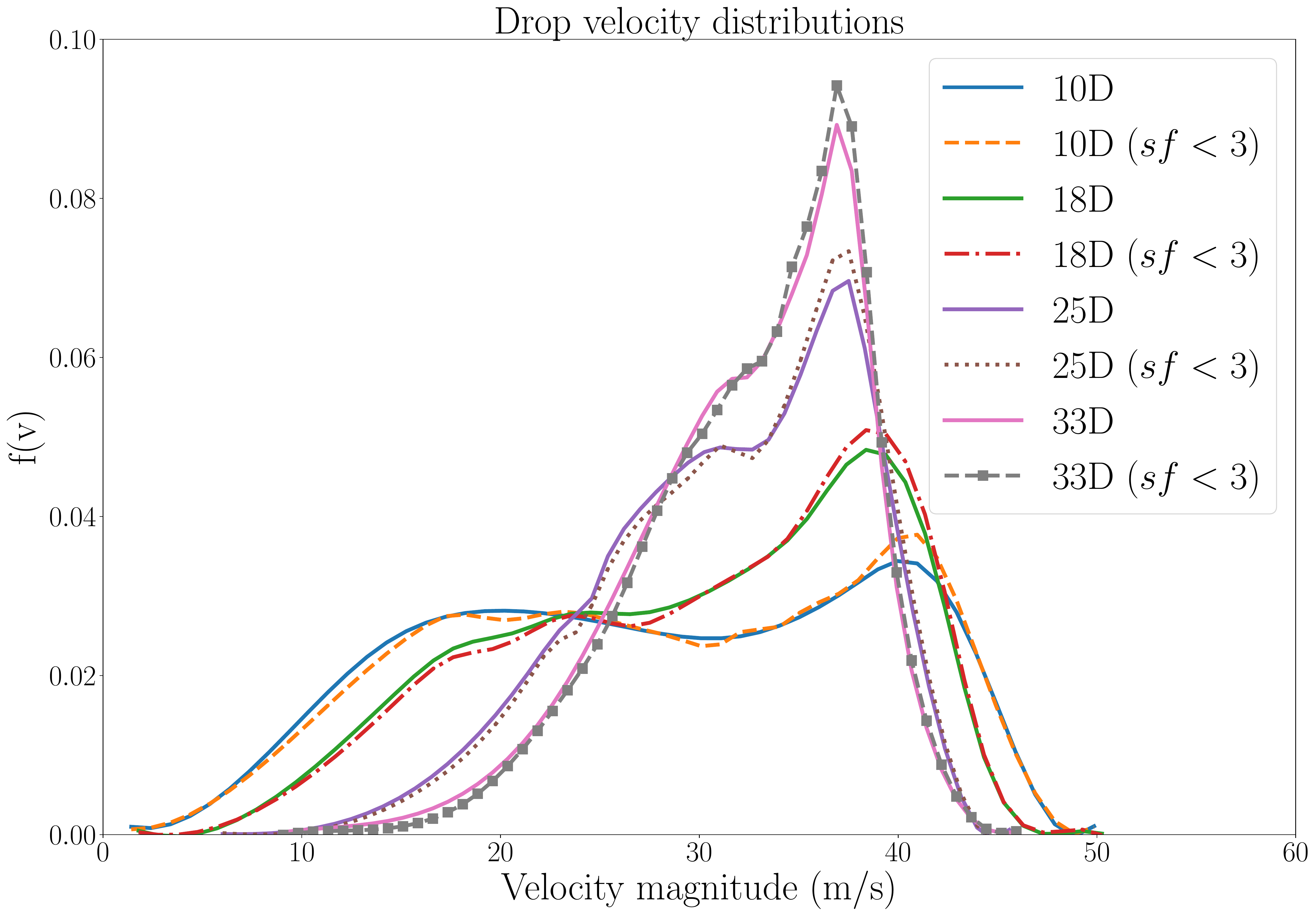}
    \caption{Comparison of probability density functions for drop size $f(d)$ and drop velocity $f(v)$ for the $q20sn0$ case at downstream locations of 10D, 18D, 25D and 33D by including all drops and by including only drops with shape-factor value less than 3.}
    \label{fig:compare_sensitivity}
\end{figure}

In Figure \ref{fig:compare_sensitivity} (left), we present the comparison of probability density function for drop size, $f(d)$ for the $q20sn0$ case at downstream locations of $10D$, $18D$, $25D$ and $33D$ by including all the drops and by including only those drops with a value of shape-factor less than 3. In Figure \ref{fig:compare_sensitivity} (right), we also present the comparison of probability density function for drop velocity, $f(v)$ for the $q20sn0$ case at downstream locations of $10D$, $18D$, $25D$ and $33D$ by including all the drops and by including only those drops with a value of shape-factor less than 3. Though, there are quantitative differences in $f(d)$ in terms of higher percentage of smaller drops than larger drops in the distribution when only the drops with shape-factor values less than 3 are considered compared to the distribution when all the drops are considered; qualitatively, both $f(d)$ and $f(v)$ follow the same behavior as the drops convect downstream from the jet. Hence the inclusion/exclusion of ligaments (elongated drops) do not significantly alter the analysis based on drop distributions for the setup presented in this work.

\section{Acknowledgements}
This project was supported by Pratt and Whitney, USA. GT would like to thank Prof. M. Herrmann (Arizona State University) for his insightful comments.

\section*{References}

\bibliographystyle{elsarticle-harv}

\bibliography{references}

\end{document}